\documentclass[fleqn,usenatbib,useAMS]{mnras}

\usepackage{graphicx}	% Including figure files
\usepackage{amsmath}	% Advanced maths commands
\usepackage{amssymb}	% Extra maths symbols
\usepackage{multicol}        % Multi-column entries in tables
\usepackage{bm}		% Bold maths symbols, including upright Greek
\usepackage{pdflscape}	% Landscape pages
\usepackage{cancel}     % strike through recommendations

\usepackage{subcaption}
\usepackage{caption}
\newcommand{\red}[1]{\textcolor{black}{#1}}

\newcommand{\ahod}{\textsc{AbacusHOD}}
\usepackage[T1]{fontenc}
\usepackage{ae,aecompl}

\usepackage{newtxtext,newtxmath}
\title[Forward Model]{Full forward model of galaxy clustering statistics with \textsc{AbacusSummit} lightcones}

\author[Yuan et al]{
Sihan Yuan,$^{1, 2}$\thanks{E-mail: sihany@stanford.edu}
 Boryana Hadzhiyska$^{3, 4}$,
 and Tom Abel$^{1, 2}$
\\
$^{1}$Kavli Institute for Particle Astrophysics and Cosmology, Stanford University, Stanford, CA 94305, USA\\
$^{2}$Department of Particle Physics and Astrophysics, SLAC National Accelerator Laboratory, Stanford, CA 94305, USA\\
$^{3}$Department of Physics, University of California, Berkeley, CA 94720\\
$^{4}$Lawrence Berkeley National Laboratory, One Cyclotron Road, Berkeley, CA 94720, USA
}

\date{Accepted XXX. Received YYY; in original form ZZZ}

\pubyear{2022}

\begin{document}
\label{firstpage}
\pagerange{\pageref{firstpage}--\pageref{lastpage}}
\maketitle

\begin{abstract}
Novel summary statistics beyond the standard 2-point correlation function (2PCF) are necessary to capture the full astrophysical and cosmological information from the small-scale ($r < 30h^{-1}$Mpc) galaxy clustering. However, the analysis of beyond-2PCF statistics on small scales is challenging because we lack the appropriate treatment of observational systematics for arbitrary summary statistics of the galaxy field. \red{In this paper, we develop a full forward modeling pipeline for a wide range of summary statistics using the large high-fidelity \textsc{AbacusSummit} lightcones that accounts for many systematic effects but also remains flexible and computationally efficient to enable posterior sampling.} We apply our forward model approach to a fully realistic mock galaxy catalog and demonstrate that we can recover unbiased constraints on the underlying galaxy--halo connection model using two separate summary statistics: the standard 2PCF and the novel $k$-th nearest neighbor ($k$NN) statistics, which are sensitive to correlation functions of all orders. \red{We will extend this method to a full cosmology emulator in a follow up paper. We expect this to become a powerful approach when applying to upcoming surveys such as DESI where we can leverage a multitude of summary statistics across a wide redshift range to maximally extract information from the non-linear scales.}
\end{abstract}


%%%%%%%%%%%%%%%%%%%%%%%%%%%%%%%%%%%%%%%%%%%%%%%%%%

%%%%%%%%%%%%%%%%% BODY OF PAPER %%%%%%%%%%%%%%%%%%

\section{Introduction}
% small scales are highly informative but requires simulation-based models for precision
The spatial distribution of galaxies presents one of the most powerful probes of the fundamental properties of the universe. Over the last few decades, galaxy clustering has emerged as an essential tool in constraining cosmology and galaxy evolution, especially with the advent of wide-field spectroscopic surveys such as the SDSS-III Baryon Oscillation Spectroscopic Survey \citep[BOSS;][]{2013Dawson}, the
SDSS-IV extended Baryon Oscillation Spectroscopic Survey \citep[eBOSS;][]{2016Dawson}, the ongoing the Dark Energy Spectroscopic Instrument \citep[DESI;][]{2013Levi, 2016DESI}, and the Prime Focus Spectrograph \citep[PFS;][]{2014Takada}. 

The standard approach to extracting cosmological information from galaxy clustering is through standard rulers and compressed statistics, most notably the baryon acoustic oscillation peak \citep[BAO;][]{2005Eisenstein} and the Alcock--Paczynski (AP) effect \citep[][]{1979Alcock} measured from the 2-point correlation function (2PCF). While the standard techniques are robust to most observational systematics \citep{2012Ross, 2015bRoss}, they are also very limited in the amount of information they can extract from the data. One key limitation is that such techniques are limited to large scales, as the theory templates rely on perturbation theories, which are only reliable beyond approximately 30--50$h^{-1}$Mpc \citep[][]{2009Carlson, 2013Carlson}. However, modern cosmological surveys are designed in a way that their galaxy clustering measurements are most accurate at scales of a few megaparsecs, far below the limits of perturbative models. The small scales are also important because they are highly sensitive to non-linear growth, and thus turn out to be significantly more constraining on cosmic growth history. The other key limit of the standard approaches is that they only rely on compressions of the 2PCF, which is in itself a compression of the full density field. Thus, to fully take advantage of the information content of modern cosmological surveys, we need to develop accurate and unbiased models for statistics beyond the 2PCF, and on scales extending deep into the non-linear regime. 

Modeling structure on small scales is challenging. Perturbative models fail because small-scale structure is dominated by high-order contributions of both the density and velocity fields, plus non-perturbative effects arising from the dynamics beyond shell crossing, i.e., formation and evolution of galaxies (or dark matter halos) and baryonic feedback. As an alternative, a new class of models have arisen in recent years leveraging large N-body simulations instead of analytical approaches. Simulations can precisely capture the non-linear evolution of dark matter density and velocity fields, given sufficient computational resources. However, N-body simulations only simulate the gravitational growth of the total matter field and need to be paired with a robust galaxy--dark matter connection model that populates galaxies on top of the simulated matter density field. A series of recent studies have attempted to obtain cosmological constraints using simulation-based models \citep[e.g.][]{2019Zhai, 2021Lange, 2021Chapman, 2021Kobayashi, 2022bYuan}. 

% high order statistics can be highly constraining but limited by systematics
At the same time, there have been numerous studies demonstrating the significant information gain when incorporating beyond-2PCF statistics. 
% \citet{2006bTinker, 2008Tinker} showed that the void probability function and under-density probability function contribute complementary information to the 2PCF due to their sensitivity to the environmental dependence of halo occupation.
% Counts-in-cells statistics have also been shown to be particularly informative on aspects of galaxy--halo connection \citep{2009Reid, 2019Walsh, 2019Wang}. 
Perhaps the most commonly discussed alternative statistics are the 3-point correlation function (3PCF) and its Fourier counterpart, the bispectrum. Several studies have successfully applied the 3PCF/bispectrum to data and obtained cosmological constraints \citep[][]{2017Slepian, 2017bSlepian, 2017Gilmarin, 2022DAmico}. However, these analyses are still limited to linear scales. \citet{2018Yuan} demonstrated the diverse information content of the squeezed 3PCF on small scales but have yet to apply it to data. 

Other extensions to the 2PCF include the marked correlation function, which weights the 2PCF with a secondary tag, such as the environment, to highlight different aspects of clustering, thus complimenting the vanilla 2PCF. A series of studies have found the marked 2PCF to potentially powerful in constraining the cosmological parameter $\sigma_8$ and modifications to general relativity \citet{2004Sheth, 2009White, 2016White}. On the small-scale front, \citet{2022Storey} forecasts the cosmological constraining power of marked statistics and void statistics in a simulation-based mock analysis and found that these statistics can bring up to $30\%$ improvements to the constraints on parameters $\sigma_8$ and $\Omega_m$.

Besides correlation-based statistics, density-based statistics such as the $k$-th nearest neighbor statistics ($k$NN) and wavelet scattering transforms (WST) have also generated considerable interest in recent years. \citet{2021Banerjee, 2021bBanerjee} found that the $k$NNs can break degeneracies in the 2-point clustering and potentially improve the cosmology constraints by a factor of a few. \citet{2022aValogiannis, 2022bValogiannis} applied WST to BOSS galaxies in a preliminary analysis of the large scales and found substantial improvement in cosmological parameter constraints. 

% we propose a simulation-based full forward model that accounts for all systematic effects and are suitable for all high order statistics down to small scales
None of these studies push their beyond-2PCF analysis to small scales on data because they lack not only a robust theory template for such scales, but also a proper treatment of the effects of observational systematics. Such systematics include: (1) redshift-dependent completeness, where the galaxy sample's number density varies as a function of redshift due to survey selection cuts; (2) survey masks and geometry, where galaxies in certain regions of the survey fail to be observed due to survey windows and various foreground obstructions; and last but not least (3) fibre collision, where galaxies that are too close to each other in projection do not always get redshift measurements. \citet{2012Ross} details these systematic effects in SDSS/BOSS, but these effects are generic to all fibre-fed spectroscopic surveys, including all ground-based experiments such as eBOSS, DESI, and PFS. 

The effects of fibre collision are particularly troublesome as they are density dependent and propagate to large scales. Traditionally, one can correct for the effects of missing redshifts through a weighting scheme that essentially assigns the weight of the missing galaxy to its nearest neighbor, as was done for BOSS \citep[][]{2014Anderson, 2016Reid}. However, while this correction works well on large scales, it introduces non-negligible spurious signal on small scales \citep[][]{2012Guo, 2006Li}. More sophisticated correction schemes have since been developed that can produce unbiased corrections for the 2PCF down to small scales \citep[e.g.][]{2020Mohammad, 2019Smith, 2018Bianchi}, but these techniques are based on recovering pair counts instead of recovering individual redshifts, thus they are only suited for the 2PCF. Because of these limitations, we lack a robust framework to extract the rich information in the non-linear scale clustering with beyond-2PCF statistics and a proper accounting of systematics. 

% redshift evolution
\red{
Another important systematic comes from ignoring redshift evolution in the observed density field. Specifically, spectroscopic surveys provide strong cosmological constraints because they access a larger number of modes in a 3D volume. However, the distribution of the matter density field is evolving as a function of redshift, and this evolution is often ignored in existing clustering analyses, leading to a potential bias. This bias will become more significant as current and future surveys push deeper in magnitude, thus probing significantly larger redshift ranges while also reaching higher measurement precision.
}

\red{In this paper, we formulate a full forward model framework that accounts for the observational systematics and redshift evolution, while also providing sufficient precision and efficiency to allow for fast and accurate model evaluations. This model framework allows for robust analyses of arbitrary galaxy clustering statistics down to highly non-linear scales. We start with the \textsc{AbacusSummit} simulations, which produce high resolution matter density field in large volumes across wide redshift epochs. We cast these simulations on redshift-evolved lightcones, generate galaxies with sophisticated galaxy--halo connection models, and add on a large range of survey systematics to produce a realistic forward model of spectroscopic survey catalogs. Our framework is highly flexible, and is particularly suited for model inferences with future spectroscopic surveys such as DESI and PFS. } 

\red{The idea of forward modeling observed galaxy field using simulated lightcones is not new \cite[e.g.][]{2022Wechsler, 2022Hahn, 2022Tam, 2018Sinha}. However, the novelty of our framework lies in the combination of the precision of the simulations used, the realism and flexibility deployed in the model, and the computational efficiency that enables fast sampling of the posterior parameter space. In terms of precision, the \textsc{AbacusSummit} lightcones we use are specifically designed to meet and exceed the needs of the new generation of spectroscopic surveys, thus our model provides both significantly higher resolution and larger volume than previous approaches. We also achieve considerable model flexibility by allowing for sophisticated and user-customised galaxy--halo connection modeling. Finally, unlike previous approaches, our forward model pipeline is highly optimised and thus allows for efficient sampling of model parameter space. 
We demonstrate these attributes of our approach by performing model parameter recovery using two sets of clustering statistics on a mock galaxy catalog that has multiple layers the systematics built in. We show that such a forward model approach is computationally efficient and achieves robust parameter constraints.}

\red{This is the first of a series of papers that will eventually extend this framework to include a full cosmology model by introducing new \textsc{AbacusSummit} lightcones at close to 100 different cosmologies, develop optimal summary statistics, and finally analyse DESI samples over a wide reshift range and derive joint constraints on cosmology and galaxy bias. We expect that further development of this approach will bring forth a zoo of analyses that finally unlock the rich information of the non-linear scales. }

This paper is structured as follows: In Section~\ref{sec:model}, we describe the tools and steps necessary in such a forward model. In Section~\ref{sec:fibre}, we have a dedicated discussion of our proposed treatment of fibre collision. In Section~\ref{sec:recovery}, we perform mock parameter recovery with with two different clustering statistics, the 2PCF and the $k$-th nearest neighbor cumulative distribution function ($k$NN-CDF). In Section~\ref{sec:discuss}, we expand on this analysis and look beyond to a full cosmology analysis on data. Finally, we conclude in Section~\ref{sec:conclude}.

\begin{figure}
    \centering
    \hspace*{-0.6cm}
    \includegraphics[width = 3.5in]{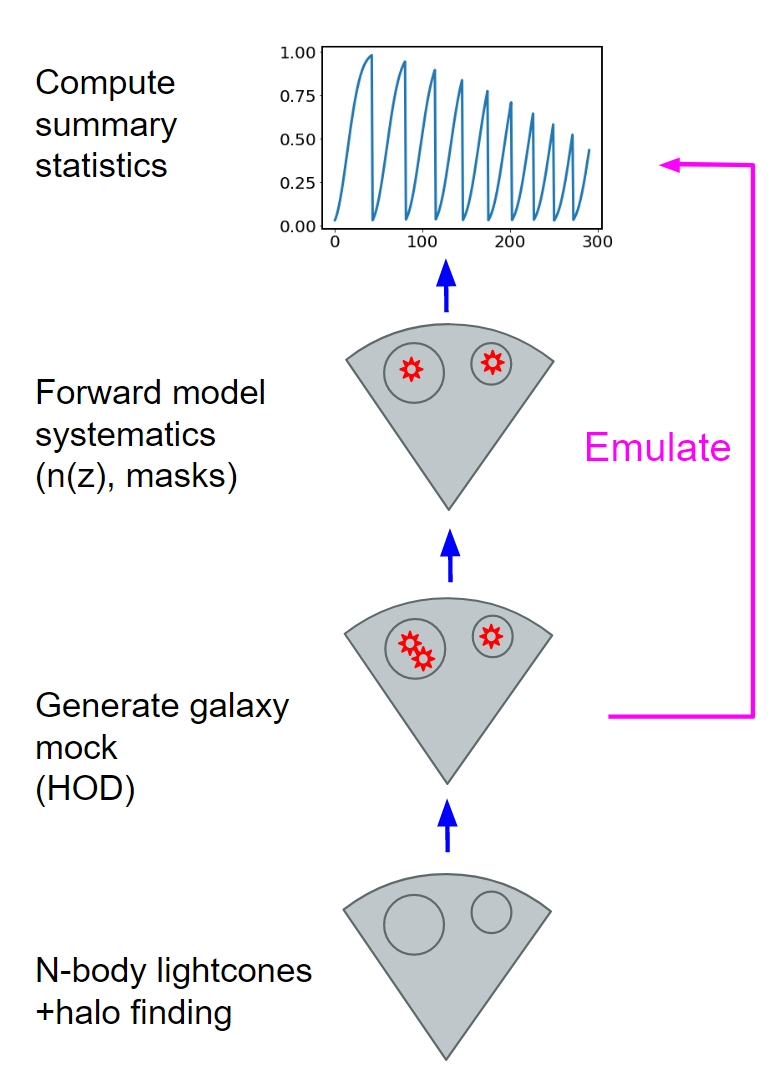}
    \vspace{-0.3cm}
    \caption{A diagram illustrating the full forward model approach using simulation lightcones. Each of the 4 forward modeling steps is described in subsections \ref{subsec:lightcones}-\ref{subsec:summarystats}, respectively. We note that fibre collision is treated separately in section~\ref{sec:fibre}.}
    \label{fig:diagram}
\end{figure}

\section{Forward model with simulation lightcones}
\label{sec:model}
In this section, we describe the forward modeling approach in detail. Figure~\ref{fig:diagram} provides a summary flowchart of the construction of the full forward model. We start with \textsc{AbacusSummit} lightcones, and we populate with mock galaxies using an HOD prescription. Then we apply layers of observational systematics including redshift selection ($n(z)$) and survey windows/masks. Finally we compute the desired summaries statistics, which can then be compared with data for likelihood analyses. We describe each of these steps in detail in the following subsections. 

\subsection{\textsc{AbacusSummit} lightcones}
\label{subsec:lightcones}

The \textsc{AbacusSummit} simulation suite \citep[][]{2021Maksimova} is a set of large, high-accuracy cosmological N-body simulations using the \textsc{Abacus} N-body code \citep{2019Garrison, 2021bGarrison}, designed to meet and exceed the Cosmological Simulation Requirements of the Dark Energy Spectroscopic Instrument (DESI) survey \citep{2013arXiv1308.0847L}. \textsc{AbacusSummit} consists of over 150 simulations, containing approximately 60 trillion particles at 97 different cosmologies. 
For this analysis, we use exclusively the ``base'' configuration boxes within the simulation suite, each of which contains $6912^3$ particles within a $(2h^{-1}$Gpc$)^3$ volume, corresponding to a particle mass of $2.1 \times 10^9 h^{-1}M_\odot$. \footnote{For more details, see \url{https://abacussummit.readthedocs.io/en/latest/abacussummit.html}}
The \textsc{AbacusSummit} suite also uses a specialised spherical-overdensity based halo finder known as {\sc CompaSO} \citep{2021Hadzhiyska}.

In addition to periodic boxes, the simulation suite also provides a set of simulation lightcones at fiducial cosmology \citep[][]{2022Hadzhiyska}. The basic algorithm associates the halos from a set of coarsely-spaced snapshots with their positions at the time of light-cone crossing by matching halo particles to on-the-fly light cone particles. The resulting halo catalogs provide accurate interpolated position for all available ``cleaned'' halos in the simulation and are particularly reliable for halos with masses above $M_\mathrm{halo} \gtrsim 1\times 10^{11}h^{-1}M_\odot$, which is more than sufficient for the purposes of current redshift surveys. Unlike other methods, which commonly adopt a ``cookie-cutting'' strategy of selecting halos at their momentary positions and thus ignore the non-negligible distance traversed by halos between redshift epochs, the lightcone catalogues of \textsc{AbacusSummit} provide interpolation that has been shown to be accurate to less than a percent \citep[for a more detailed discussion, see Section 4 in][]{2022Hadzhiyska}. For this analysis, we utilise the 25 base lightcones, which are constructed from the 25 base periodic boxes, with each lightcone covering an octant of the sky ($\sim 5156$ deg$^2$) up to $z\sim 0.8$. 

\red{Throughout this section, we use CMASS as an example, but the techniques we describe are generic to any spectroscopic galaxy survey.} The CMASS sample is approximately 9000 deg$^2$ in area and spans redshift range $0.45<z<0.6$. Thus, we can exceed the data volume with two base lightcones, though ideally we want to use even more lightcones to further reduce model sample variance. 

\subsection{\textsc{AbacusHOD} on lightcone}
The first step of the forward model is to populate the simulation lightcones with galaxies. To achieve this, we use a Halo Occupation Distribution \citep[HOD; e.g.][]{2005Zheng, 2007bZheng} approach, which probabilistically populate dark matter halos with galaxies according to a set of halo properties. For a Luminous Red Galaxy (LRG) sample, the HOD is well approximated by a vanilla model given by (originally shown in \citet{2015Kwan}):
\begin{align}
    \bar{n}_{\mathrm{cent}}^{\mathrm{LRG}}(M) & = \frac{\mathrm{ic}}{2}\mathrm{erfc} \left[\frac{\log_{10}(M_{\mathrm{cut}}/M)}{\sqrt{2}\sigma}\right], \label{equ:zheng_hod_cent}\\
    \bar{n}_{\mathrm{sat}}^{\mathrm{LRG}}(M) & = \left[\frac{M-\kappa M_{\mathrm{cut}}}{M_1}\right]^{\alpha}\bar{n}_{\mathrm{cent}}^{\mathrm{LRG}}(M),
    \label{equ:zheng_hod_sat}
\end{align}
where the five vanilla parameters characterizing the model are $M_{\mathrm{cut}}, M_1, \sigma, \alpha, \kappa$. $M_{\mathrm{cut}}$ characterises the minimum halo mass to host a central galaxy. $M_1$ characterises the typical halo mass that hosts one satellite galaxy. $\sigma$ describes the steepness of the transition from 0 to 1 in the number of central galaxies. $\alpha$ is the power law index on the number of satellite galaxies. $\kappa M_\mathrm{cut}$ gives the minimum halo mass to host a satellite galaxy.
\red{We have added a modulation term $\bar{n}_{\mathrm{cent}}^{\mathrm{LRG}}(M)$ to the satellite occupation function to remove the possibility of populating satellite galaxies in small and poorly resolved halos. However, there is evidence that such central-less satellites may exist in a realistic stellar-mass selected catalog \citep[][]{2019Jimenez}. }
% In the description of their HOD model, the authors say: “ We have added a modulation term n_cen (M) to the satellite occupation function to remove satellites from halos without centrals.” Using a HOD formalism in conjunction with the "halo model" necessitated this approximation to remove some terms from the final equations, but modern HOD run on dark matter simulations eliminates this requirement. In addition, works such as Jimenez et al. (2019) demonstrated that in galaxy formation models it is possible to have haloes only occupy by satellite galaxies (when selecting the galaxies by stellar mass). Even further, Chavez-Montero et al. (2022) demonstrated that ignoring this effect will reduce the HOD model's ability to accurately reproduce galaxy-galaxy lensing. I would request that the authors mention the limitations of this approximation. If the author has a compelling argument for why this term should be included, I will request that they discuss it in the text.

We have also included an incompleteness parameter $\mathrm{ic}$, which is a downsampling factor controlling the overall number density of the mock galaxies. \red{This parameter is conceived to account for incompleteness in the observed galaxy sample and is tuned by matching against the mean number density in the observed sample \citep[e.g.][]{2016Rodriguez, 2016Leauthaud, 2018Guo}.} By definition, $0 < \mathrm{ic}\leq 1$.

In addition to determining the number of galaxies per halo, the standard HOD model also dictates the position of velocity of the galaxies. For the central galaxy, its position and velocity are set to be the same as those the halo center, specifically the L2 subhalo center-of-mass for the {\sc CompaSO} halos. For the satellite galaxies, they are randomly assigned to halo particles with uniform weights, each satellite inheriting the position and velocity of its host particle. 

For this paper, we fix two parameters $\sigma$ and $\kappa$ in the vanilla HOD for simplicity. $\kappa$ does not strongly affect clustering and only comes into effect at very small scales. $\sigma$ does affect clustering on 2-halo scales, but it tends to be strongly degenerate with $\log M_\mathrm{cut}$. We omit $\sigma$ in this preliminary analysis for clearer interpretation of the results. We also ignore redshift-dependence in the HOD, setting $\mu_{\rm cut, p} = 0$ and $\mu_{1, p} = 0$. Thus, in the following analysis, the HOD is fully parameterised by 4 parameters, $M_{\mathrm{cut}}, M_1, \alpha$, and $\mathrm{ic}$.

In order to sample the model parameter space, each forward model step needs to be computationally efficient so as to minimise the time need to evaluate the full forward model. To this end, we adopt the highly optimised \ahod\ implementation, which significantly speeds up the HOD calculation per HOD parameter combination \citep[][]{2021bYuan}. The code also enables a range of physically motivated extensions to the vanilla HOD and also redshift-dependent HODs (see Appendix~\ref{a:hodz}). The code is publicly available as a part of the \textsc{abacusutils} package at \url{http://https://github.com/abacusorg/abacusutils}. Example usage can be found at \url{https://abacusutils.readthedocs.io/en/latest/hod.html}.

% For this analysis, we invoke a set of HOD extensions known as velocity bias, which biases the velocities of the central and satellite galaxies. This is shown to to be a necessary ingredient in modeling BOSS LRG redshift-space clustering on small scales \citep[e.g.][]{2015aGuo, 2021bYuan}. Velocity bias has also been identified in hydrodynamical simulations and measured to be consistent with observational constraints \citep[e.g.][]{2022Yuan, 2017Ye}. We do not invoke any galaxy assembly bias in this analysis for simplicity, but we fully expect our conclusions to extend to more complex HOD models. 

% We parametrise velocity bias through two additional HOD parameters: \texttt{$\alpha_\mathrm{vel, c}$} is the central velocity bias parameter, which modulates the peculiar velocity of the central galaxy relative to the halo center. $\alpha_\mathrm{vel, c} = 0$ indicates no central velocity bias, i.e. centrals perfectly track the velocity of halo centers. 
% \texttt{$\alpha_\mathrm{vel, s}$} is the satellite velocity bias parameter, which modulates how the satellite galaxy peculiar velocity deviates from that of the local dark matter particle. $\alpha_\mathrm{vel, s} = 1$ indicates no satellite velocity bias, i.e. satellites perfectly track the velocity of their underlying particles.  

\subsection{Survey systematics}
\label{subsec:systematics}

In our forward model, we account for both redshift-dependent completeness and the survey geometry. Both of these systematic effects can significantly bias the measurements but are hard to model from a periodic simulation box. Simulation lightcones allow these effects to be modeled relatively straight-forwardly. A third critical systematic effect is fibre collision, but we reserve that discussion for Section~\ref{sec:fibre}.

To apply redshift-dependent completeness, we compute the density of galaxies generated by the HOD $n_\mathrm{HOD}$, and then run a filtering step where we retain each galaxy with probability $p(z) = n_\mathrm{data}(z)/n_\mathrm{HOD}$. This step ensures the resulting number density profile mimics the observation $n_\mathrm{data}(z)$. This step is computationally efficient and can be trivially parallelised. 

Due to the complex geometry of the survey boundaries and masks, any summary statistics measured on a realistic sample suffers from boundary effects. To model such effects in a forward model, ideally one would want to generate a sufficiently large lightcone that would enclose the entire survey footprint. Then, one can simply account for such boundary effects by imposing the survey boundaries and masks on the lightcone mock. However, each of our lightcones is only an octant of the sky, approximately half of the CMASS footprint. Thus, we cannot directly model the entire set of survey boundaries. 
As a compromise, we trim the data and the lightcone to share identical geometry, at the cost of throwing away a fraction of the data. 

\begin{figure}
    \centering
    \hspace*{-0.6cm}
    \includegraphics[width = 3.7in]{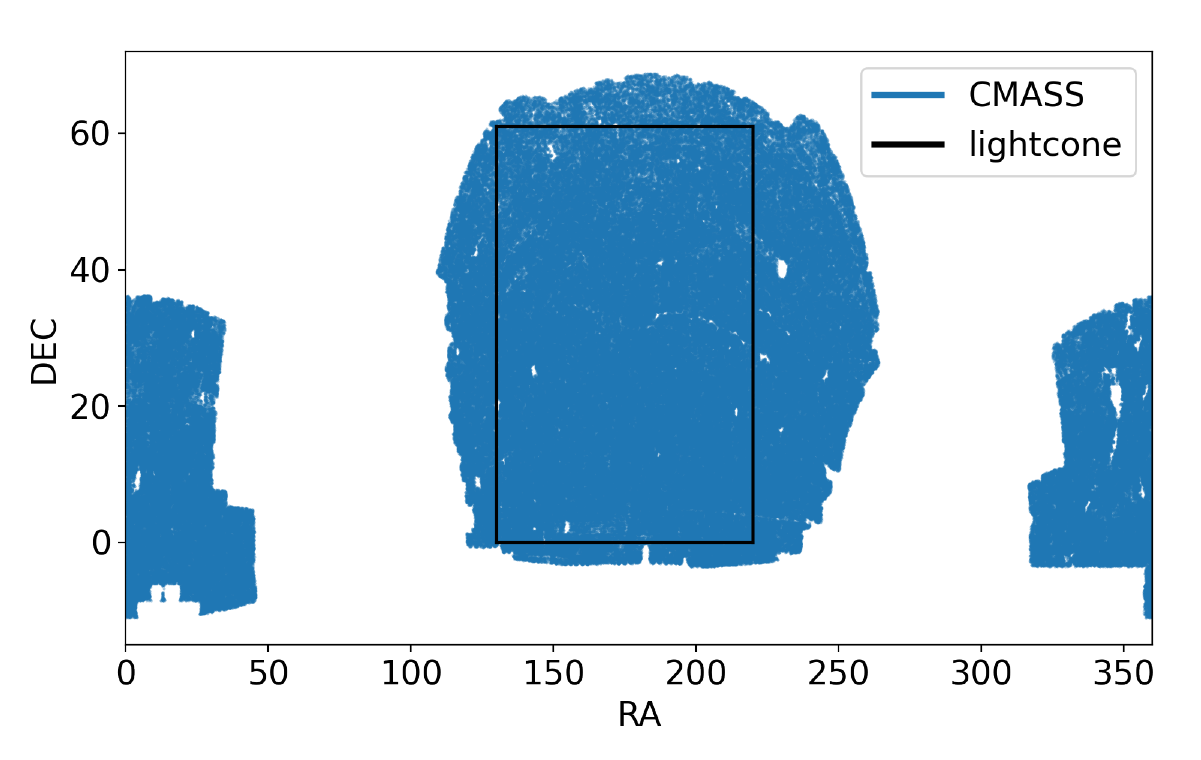}
    \vspace{-0.3cm}
    \caption{The CMASS LRG footprint (indicated with blue points) compared to the trimmed footprint (enclosed by black lines). The lightcone footprint is rotated in RA to have maximally continuous overlap with the CMASS footprint. We apply a cut at DEC = 61$^\circ$ to remove regions of the lightcone that do not overlap with the CMASS footprint.}
    \label{fig:footprint}
\end{figure}

Using the CMASS sample as an example, we start by rotating the lightcone coordinates to maximally overlap with the CMASS footprint, as illustrated in Figure~\ref{fig:footprint}. In this case, the rotation is only along the RA direction and results in the lightcone spanning 130-220$^\circ$ in RA. 
Then we apply a cut at DEC = 61$^\circ$ to remove regions of the lightcone that do not overlap with the CMASS footprint. With this cut, the resulting footprint is fully enclosed in the CMASS survey footprint. We propose to trim both the data and the lightcone to this ``rectangular'' footprint to guarantee the the model and the data have identify boundary effects. 

Finally, we apply the additional survey masks in the model, including the bright star mask, which accounts for the footprints missing due to bright forground stars, the centerpost mask, which accounts for holes in the footprint due to the centerpost of tiles, and masks accounting for bad tiles. 

Last but not least, the forward model also needs to account for effects of fibre collisions, the effect where spectroscopic fibres can not be placed infinitely close to each other, resulting in missing galaxies in dense regions. We reserve the treatment of fibre collision to a dedicated discussion in Section~\ref{sec:fibre}. In summary, instead of forward modeling fibre collision, which is computationally expensive, we construct a routine to probabilistically recover the redshifts of collided galaxies and demonstrate that the resulting systematic error is subdominant compared the sample variance. 

\subsection{Summary statistics}
\label{subsec:summarystats}

Having generated the lightcone galaxy catalogs and forward-modeled the full range of systematics, one can now compute the desired summary statistics and perform likelihood analysis against the data. While our approach is fully applicable to any set of clustering statistics, we focus on the novel $k$-th nearest neighbor statistics and the more widely used 2-point correlation function in this paper. We focus on $k$NNs because they are  computationally efficient ($\mathcal{O}(N\log N)$) while incorporating high-order clustering information. However, because they are directly based on density distributions, $k$NNs are also highly sensitive to systematics that result in missing objects. Such sensitivities highlight the need for a full forward model approach. 

\subsubsection{$k$-th nearest neighbor statistics}
\label{subsubsec:knn}
In this section, we give a quick review of the $k$NN formalism and our Python implementation. For detailed derivations and illustrations, we refer the readers to \citet{2021Banerjee, 2021bBanerjee}. To define $k$NNs, we first define $P_{k|V}$, the probability of finding exactly $k$ data points in a volume $V$, averaged over all query points inside the survey volume.
We can write out $P_{k|V}$ in terms of its cumulative counterparts as
\begin{equation}
    \label{eq:3}
    P_{k|V} = P_{>k-1|V} - P_{>k|V},\ \mathrm{for}\ \forall k\geqslant 1\,,
\end{equation}
We can define cumulative distribution functions as
\begin{equation}
    \mathrm{CDF}_{(k+1)\mathrm{NN}}(r) = P_{>k|V=\frac{4\pi}{3} r^3} = 1 - \sum_{i = 0}^{k} P_{i|V=\frac{4\pi}{3} r^3},
    \label{equ:cdf}
\end{equation}
where we have also reformulated the CDFs as a function of radial distance between the query point and the data point $r$. 
These CDFs as a function of $r$ and of order $k$ form a series of summary statistics that we later refer to as the $k$NNs. \citet{2021Banerjee} showed that the $k$NNs automatically includes information from all orders of correlation function without the penalty of increased functional complexity. For plotting purposes, we also define the peaked CDFs (pCDF) as 
\begin{align}
\mathrm{pCDF}_{k}(r) = \begin{cases} \textrm{$k$NN-CDF}(r) & \text{if $\textrm{$k$NN-CDF}(r) < 0.5$}, \\
1-\textrm{$k$NN-CDF}(r) & \text{otherwise}. \end{cases}
\end{align}
Conceptually, one can think of the $k$NN-CDF as the cumulative distribution of the distances from query points to the $k$-th nearest neighbors. 

The calculation of the $k$NN-CDF is straightforward. We first generate a large grid of query points. For this paper, we adopt a grid spacing of $l = 4h^{-1}$Mpc and then only select the grid points that are within the trimmed survey volume we consider. Then we construct a $k$DTree of the 3D positions of galaxies, from which we inquire the distance to the $k$-th nearest neighbor of all the query points. Finally, we sort the distances from all the query points and construct a CDF. 

It is worth noting that the choice of $l = 4h^{-1}$Mpc effectively imposes a minimum scale for the $k$NN-CDF, as scales below grid spacing $l$ will be poorly sampled and thus carry significant noise. This point becomes important when we set up the mock test in section~\ref{sec:recovery}. In principle, reducing the grid spacing would allow us to access $k$NNs on smaller scales, but at significant computational and memory cost. We argue that this coarse spacing is sufficient for the purpose of this pilot study, but advocate for more advanced techniques for more optimal sampling of query points, such as the one proposed in Appendix~A of \citet{2022Garrison}.

\subsubsection{2-point correlation function}
We compare the $k$NN-CDF with the more commonly used 2-point correlation function in this paper. Specifically, we use projected 2PCF $w_p$:
\begin{equation}
w_p(r_p) = 2\int_0^{\pi_{\mathrm{max}}} \xi(r_p, r_\pi)d\pi,
\label{equ:wp_def}
\end{equation}
where $r_p$ and $r_\pi$ are the transverse and line-of-sight (LoS) separations in comoving units. $\xi(r_p, r_\pi)$ is the redshift-space 2PCF, which can be computed via the \citet{1993Landy} estimator:
\begin{equation}
    \xi(r_p, r_\pi) = \frac{DD - 2DR + RR}{RR},
    \label{equ:xi_def}
\end{equation}
where $DD$, $DR$, and $RR$ are the normalised numbers of data-data, data-random, and random-random pair counts in each bin of $(r_p, r_\pi)$.  For implementation, we use the highly-optimised grid-based \textsc{Corrfunc} code \citep{2020Sinha} for fast calculations. In the case of the lightcones, which have a more complex geometry: namely, three boxes intersected by concentric shells \citep[see Fig. 1 in][]{2022Hadzhiyska}, we generate randoms by populating an octant of a shell of thickness determined by the lightcone crossing comoving distance for each redshift epoch, disposing of particles outside the three boxes at higher redshifts.
$w_p$ is commonly used in cosmology because it marginalises over the LoS positions of galaxies, which tend to suffer from significant redshift uncertainties, especially in the case of photometry-only data. However, the marginalisation comes at the cost of losing out on the information embedded in the LoS structure of $\xi(r_p, r_\pi)$. In Section~\ref{sec:fibre}, we discuss one source of such redshift uncertainty and an effective remedy for it. 

\subsection{Computational efficiency}

\red{A key requirement of our forward model is that it needs to be not only realistic, but also computationally efficient in order to enable sampling of the posterior parameter space. 
This will become particularly important when we enable cosmology sampling via an emulator of \textsc{AbacusSummit} lightcones at different cosmologies in a future paper. Such an emulator analysis would require constructing forward models and sampling HOD posteriors on approximately 100 lightcones. To this end, it is essential to characterise and optimise the computational efficiency of each step of the forward model at this stage. }

In this section, we report the timing and computational speed-ups we implemented for our forward modeling steps. The timing is done on a modest machine with two Intel Xeon Gold 5218 chips clocked at 2.3 GHz for a total of 32 physical cores and 256 GB DDR4-2666 RAM. 

\red{
We start with generating mock galaxies on the \textsc{AbacusSummit} lightcone. The halo lightcones of \textsc{AbacusSummit} are organised in the same format as the halo catalogues of \textsc{AbacusSummit} cubic boxes, which allows us to easily transfer the optimised \ahod\ to run on the lightcones. Generating a CMASS-like LRG sample on a single lightcone covering an octant of the sky takes $\sim 0.06$ seconds. This is relatively insignificant compared to later steps modeling systematics and computing summary statistics. }

\begin{figure*}
    \centering
    \hspace*{-0.6cm}
    \includegraphics[width = 7in]{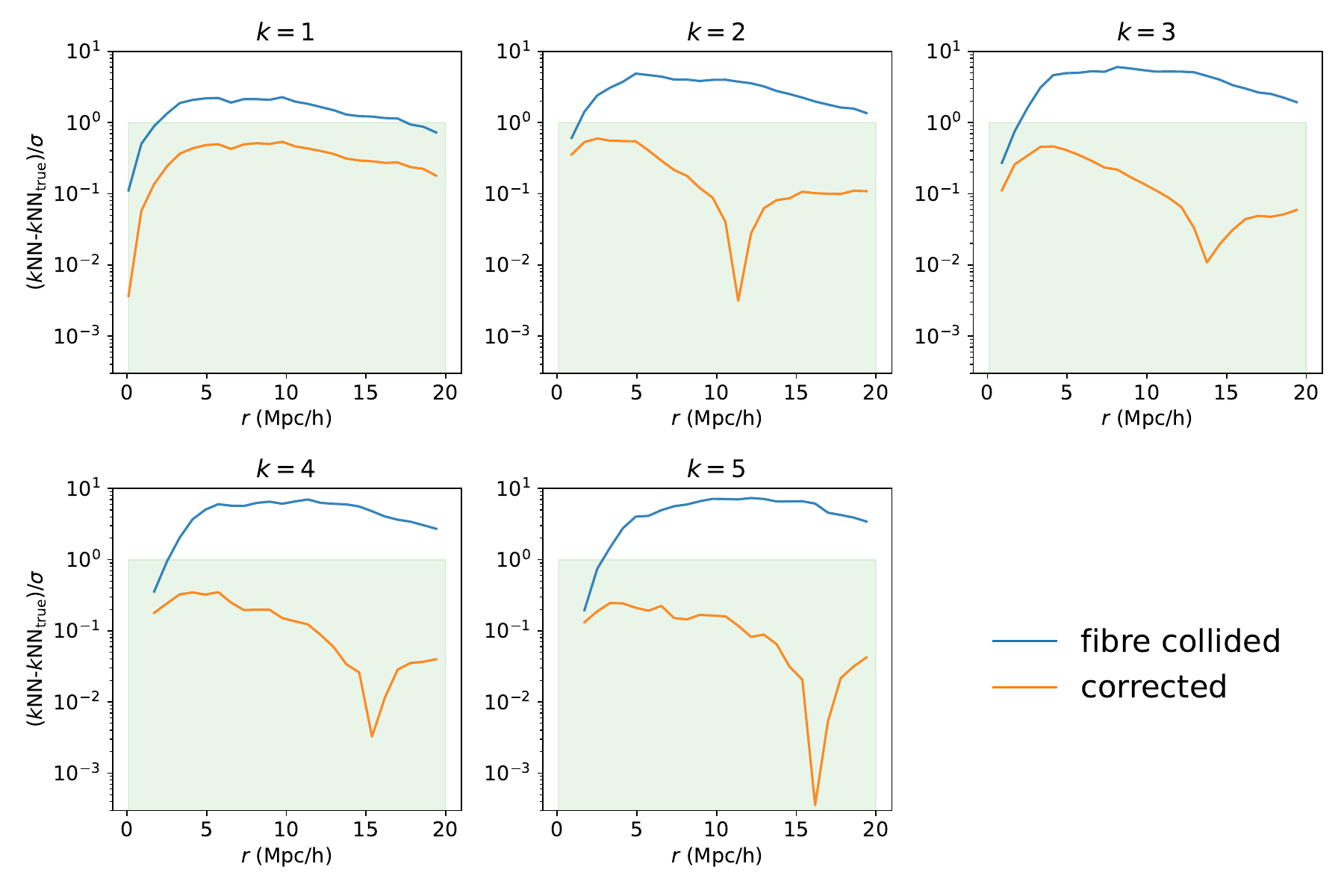}
    \vspace{-0.3cm}
    \caption{The effect of fibre collision on the $k$NN-CDF, and the recovery when applying corrections. \red{The $y$-axis represents the fractional difference between the CDF and the ``true'' pre-collision CDF, normalised by the expected sample variance in a CMASS volume. The value 0 indicates perfect agreement between the shown CDF and the ``true'' CDF. The green bands indicate the regions where the fibre collision error is less than the expected sample variance.} The blue curves show the relative difference between the CDF after applying fibre collision and the ``true'' CDF, demonstrating that the fibre collisions have a significant impact on the measured $k$NN-CDF. The orange curves indicate the CDFs once corrections have been applied to the fibre-collided catalogs, showcasing good recovery of the underlying ``true'' $k$NN-CDF measurement.}
    \label{fig:fc}
\end{figure*}

Applying the $n(z)$ filter can be trivially parallelised and each evaluation takes $\sim 0.04$ seconds across 32 cores. Applying survey mask is significantly slower at $\sim 0.5$ seconds per evaluation. The performance is bottlenecked by the existing Python implementations of \textsc{Mangle}\footnote{\url{https://github.com/esheldon/pymangle}} \citep[][]{2008Swanson, 2004Hamilton}, which is necessary to manipulate the archival BOSS survey mask files. While the application of survey masks to the galaxy catalogs can be parallelised, the run time is dominated by overhead at above $\sim 8$ threads. We note that this performance is highly specific to the BOSS survey mask files, which were developed on outdated methodologies and not optimised for performance. For upcoming datasets such as DESI, we can likely optimise the mask file formats and application algorithms for parallelisation.

Finally, the summary statistics calculation takes $\sim 0.5$ seconds for each $k$NN-CDF evaluation ($\sim 0.2$ seconds for each 2PCF evaluation). The rough breakdown of the time spent per $k$NN-CDF evaluation is the following: 1) $\sim 0.1$ seconds on transforming spherical coordinates to cartesian coordinates before constructing kDTree; 2) $\sim 0.05$ seconds on constructing kDTree; 3) $\sim 0.15$ seconds on neighbor queries; 4) $\sim 0.2$ seconds on sorting and constructing CDFs. The only step that is currently fully parallelised is step 3, and it scales well with number of cores. Step 1 can be parallelised in principle for modest performance gains. Step 2 is an intrinsically serial task, and we have not come across a successful parallel implementation, but it is a relatively cheap process as it is. Step 4 can be parallelised in principle with parallel mergesort algorithms, but our tests showed insignificant performance gains when using available parallel sorting algorithms compared to \textsc{Numpy} Quicksort. 

While the $k$NN-CDF calculation is relatively slow, it does scale well with number of radial bins (only affects step 4) and order $k$ (only affects step 3,4). Thus, $k$NN-CDF is computationally advantageous when compared to high-order correlation functions. We also expect to achieve significant speedups by building up a grid-based $k$NN calculator from scratch, adopting many of the techniques used for \textsc{Corrfunc}.

\section{fibre collision correction}
\label{sec:fibre}

Fibre collision refers to the effect where spectroscopic fibres are not infinitely thin so one cannot put two fibres infinitely close to each other. For example, in BOSS, the minimum angular distance between two fibres, known as fibre collision radius, is 62$^{\prime\prime}$. Because fibre collision is more common in over-dense regions, its effect correlates strongly with the underlying clustering. Thus, it is a important observational systematic that needs to be addressed and mitigated. In principle, one can overcome this issue by repeatedly visiting the same area of the sky with the telescope, but that significantly reduces the survey efficiency. Thus, a typical survey strategy, such as the ones for BOSS and DESI \citep[][]{2003Blanton, 2022Abareshi}, will only produce spectra for 80-98$\%$ of the targets. While the incompleteness is small, it can still produce significant effects on the measured clustering, especially at the precision achievable with DESI \citep[e.g.][]{2017Pinol, 2017Hahn}. 

Several techniques have been developed to correct for fibre collision, such as \citet{2012Guo} and \citet{2017Bianchi}. However, these techniques appeal to properties of 2-point correlation function and are not applicable for arbitrary summary statistics.

In principle, one can forward model the effects of fibre collision given the fibre assignment strategy is publicly available. However, fibre assignment codes involve computational expensive steps such as group finding and neighbor searches. These operations make the fibre assignment code prohibitively expensive to apply in repeated forward model evaluation. Thus, in this section, we demonstrate a novel approach to minimise the effect of fibre collision by applying corrections to the data on the catalog level.

The basic idea is to infer the redshift of the missing galaxies to the best of our abilities. We assume we have accurate photometric measurement of the angular positions of all the missing galaxies, and we also assume we have a corrected full-shape 2PCF measurement down to very small scales, specifically $\xi(r_p, \pi)$. Both of these assumptions are reasonable for current and upcoming spectroscopic surveys as they are often preceded by a photometric surveys for target selection and that the effects of fibre collision can be removed in the 2PCF with the previously mentioned techniques. 

\red{The key insight is that the 2PCF is essentially the probability distribution of the positions of neighboring galaxies around any arbitrary galaxy. Specifically given the transverse separation distance $r_p$, the redshift-space 2PCF $\xi(r_\pi| r_p)$ gives a one dimensional PDF for the LoS separation distance $r_\pi$. 
Thus, for each missing galaxy, we identify its $N$ closest neighbors in projected $r_p$ plane. We then use the transverse separation $r_p$ of the missing galaxy to its neighbors to sample $\xi(r_\pi| r_p)$ to statistically infer the LoS position of the missing galaxy. We can also fold in the $r_\pi$ PDFs from $N > 1$ neighbors to improve the constraints on the missing galaxy's position.} In principle, this idea is similar to ``clustering-based redshifts'' developed in \citet{2013Menard}, except this general approach is particularly suited for the fibre collision problem because every collided galaxy is necessarily close in projection to a target with known redshift. 

\begin{figure}
    \centering
    \hspace*{-0.5cm}
    \includegraphics[width = 3.5in]{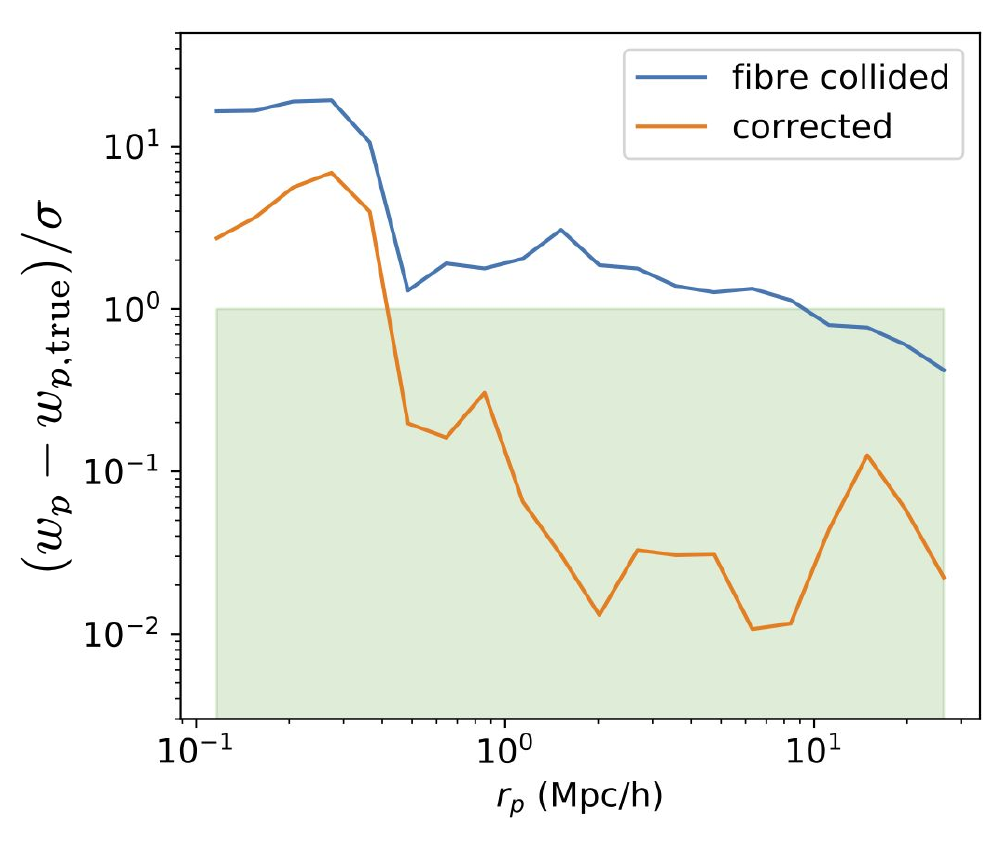}
    \vspace{-0.2cm}
    \caption{The effect of fibre collision on the projected 2-point correlation function $w_p$, and the recovery when applying corrections. The $y$-axis represents the fractional difference compared to the ``true'' pre-collision 2PCF, normalised by the expected sample variance in a CMASS volume. The green bands indicate said expected sample variance. The blue curves show the relative difference between the $w_p$ after applying fibre collision and the ``true'' $w_p$, demonstrating that the fibre collisions have a significant $\sim 5\%$ impact on the measured $w_p$, extending up to large scales. The orange curves indicate the $w_p$ once corrections have been applied to the fibre-collided catalogs, showcasing excellent recovery of the true signal down to the fibre collision radius of $\sim 0.5h^{-1}$Mpc.}
    \label{fig:fc_wp}
\end{figure}

To test the performance of this technique, we populate the 25 base simulation cones with an HOD matched to CMASS LRGs (see section~\ref{subsec:mock}), and apply the full set of selections and masks as described in section~\ref{subsec:systematics}, from which we measure the desired galaxy clustering statistics and calculate the average over 25 lightcones as the pre-collision ``true'' measurement. Then we run a BOSS-like tiling and fibre assignment code on the lightcone catalogs to separate galaxies into ones with assigned fibres and ones without. We can then measure the desired clustering statistics on the galaxies with assigned fibres and assess the effect of fibre collision. We refer the readers to \citet{2003Blanton} for a pedagogical description of the tiling and fibre assignment procedure, but offer a brief summary as follows. 

The procedure first applies tiling by drawing overlapping circle tiles around a grid of tile centers on the 2D mock sky, with tile center separation and tile radius set to BOSS values. Then within each tiled region, we run a group finder on the projected 2D galaxy field with linking length set to BOSS fibre collision distance $62''$. For each group, we identify the maximum un-collided set, which is the maximum subset of galaxies in the group that are all separated by at least $62''$. These galaxies are guaranteed fibres. Then the ones not in the maximum subset are then passed through an additional filtering step, which determines its probability of receiving a fibre based on number of tile overlaps at the location. Specifically, we base these probabilities from BOSS, $0\%$ if there is only one tile, $60\%$ if there are two tiles, $90\%$ if there are three tiles \citep[][]{2016Reid}. As a result, $\sim 5\%$ of the galaxies do not receive a fibre, consistent with the collided fraction seen in BOSS CMASS sample \citep[][]{2012Anderson, 2012Guo, 2016Reid}. 

\red{Figure~\ref{fig:fc} showcases the effect of fibre collision on the $k$NN-CDF of the lightcone mocks, averaged over 25 lightcones. The blue curves show the fractional error induced by fibre collision, normalised by the expected sample variance of a CMASS sample. The green bands represent the sample variance to aid visualisation.} Clearly, fibre collision has a significant effect on the measured signal. We repeat the same experiment for the projected 2-point correlation function $w_p$ in Figure~\ref{fig:fc_wp}, where the blue curve shows that fibre collision has a significant effect on $w_p$ extending to $r_p\sim 10h^{-1}$Mpc. The fibre collision effect on $w_p$ is approximately $5\%$ of the total $w_p$ amplitude, consistent with those reported for BOSS in \citet{2012Anderson, 2012Guo}.  

Finally we apply the correction scheme as described to recover the redshifts of the fibre collided galaxies, and measure the summary statistics on the corrected galaxy mocks.
On the catalog level, the median absolute recovery error on the LoS coordinate of the collided galaxies is $\Delta z = 8h^{-1}$Mpc.
% with a long thin tail of outliers with large $\Delta z$. The outliers are expected as the correction scheme would tend to relocate chance collisions (un-related galaxies overlapping by chance) and move them close to each other. 
If we sample the corrected position from the probabilities conditioned on the nearest two neighbors, instead of just the nearest neighbor, we get an even better recovery, with a median $z$ error of $\Delta z = 6h^{-1}$Mpc. 
%However, the outlier tails remain, and the resulting improvements to the corrected summary statistics are marginal. 

More importantly, we assess the performance of the correction on the summary statistics, starting with the $k$NN-CDFs. 
On Figure~\ref{fig:fc}, the orange curves show the corrected $k$NN-CDFs relative to the truth, where the LoS positions are inferred from just the nearest neighbor. For the $k$NN-CDFs, we can see that the correction significantly reduces the error due to fibre collision to within $1\sigma$. We also find additional improvements to the performance of the correction when including two nearest neighbors instead of just the nearest neighbor. However, given the current level of sample variance, using just the nearest neighbor is sufficient for the $k$NN-CDFs. \red{The mean residual error due to fibre collision after applying the correction is approximately $20\%$ of the CMASS sample variance uncertainty, which translates to a small $4\%$ increase to the final covariance matrix. We ignore this term in the following analysis in section~\ref{sec:recovery}, but we note that this term can become important when applying such techniques to upcoming surveys like DESI, where the effective volume is $\sim 10$ times that of CMASS \citep{2016DESI}. } 

For the 2PCF, we also find excellent recovery of the underlying true signal. 
The orange curve of Figure~\ref{fig:fc_wp} shows the performance of the correction on the projected 2-point correlation function. At scales greater than the fibre collision scale $r_p > 0.5h^{-1}$Mpc, the correction almost perfectly recovers the true signal. \red{The residual systematic error is insignificant at approximately $7\%$ of the sample variance.} At smaller scales, the scheme still results in large improvements compared to the uncorrected measurement, but the residual relative to the true signal is still significant. However, this is not a significant issue since scales below $0.4h^{-1}$Mpc are also systematics dominated and remain largely uninformative for cosmological analysis \citep[][]{2022bYuan, 2021Lange}. 
However, the success of the correction scheme on the projected 2PCF on larger scales is expected because, by definition, the projected 2PCF marginalises over the LoS positions and is thus not strongly sensitive to fibre collision effects. The large-scale effects we see for the blue curve in Figure~\ref{fig:fc_wp} is coming from the finite LoS integration length $\pi_\mathrm{max}$, which we set to $30h^{-1}$Mpc for this analysis.

To summarise, we have shown through these tests that we can successfully remove the effects of fibre collision with our redshift recovery scheme, at least to the level of precision required for a CMASS analysis on relevant scales. Further improvements to the method are likely needed for a future DESI analysis, where will utilise measurements of significantly higher precision and extending down to much smaller scales. We reserve that discussion for a future paper. 

\begin{figure}
    \centering
    \hspace*{-0.6cm}
    \includegraphics[width = 3.9in]{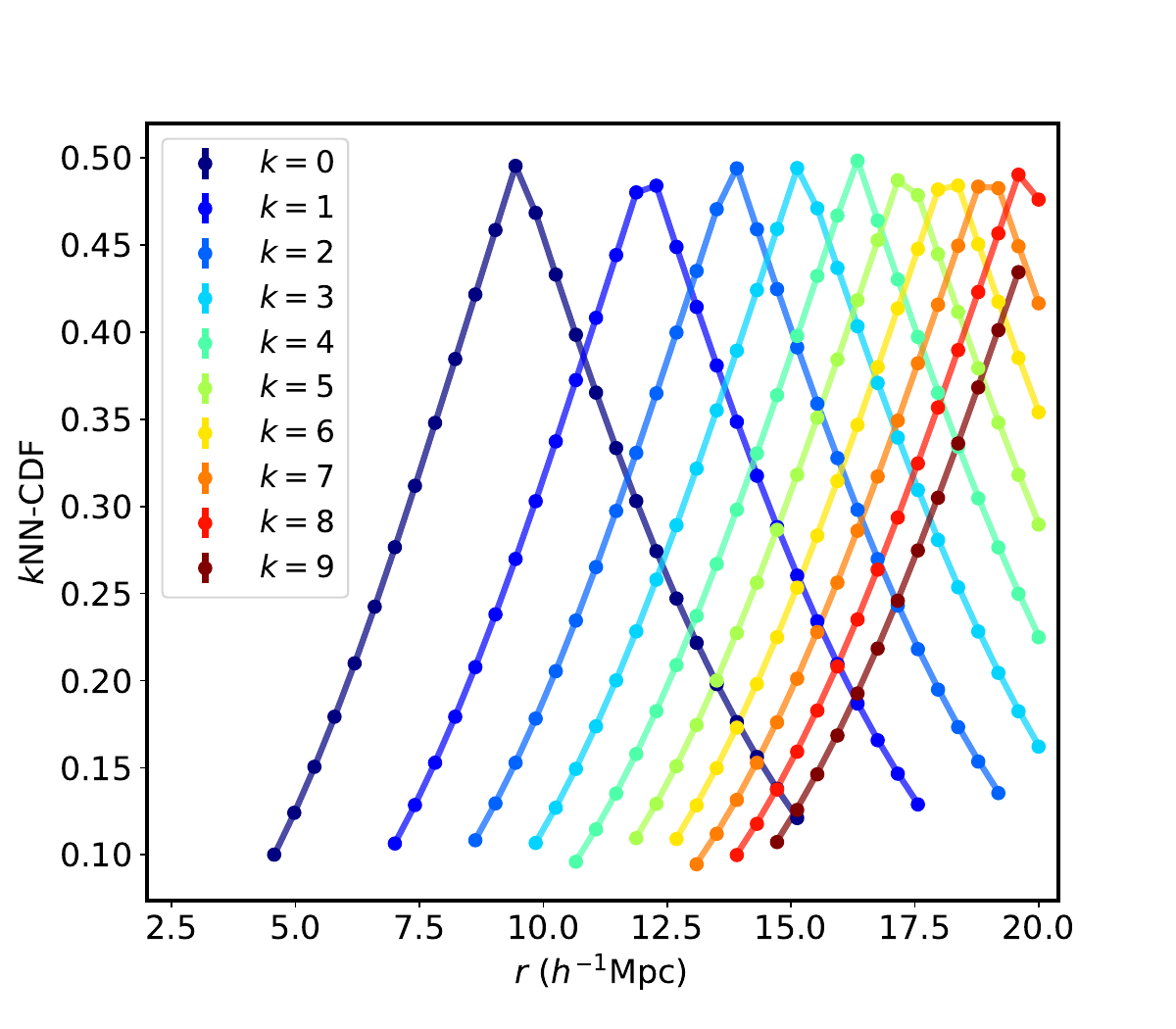}
    \vspace{-0.4cm}
    \caption{\red{The difference between the best-fit $k$NN-CDF and the mock data vector. The $y$-axis shows the relative difference between the best fit and the mock data, normalised by the CMASS error bar. The green band indicate the $1\sigma$ region. Different colors correspond to different $k$s.}}
    \label{fig:knn_target}
\end{figure}

% \begin{figure*}
%     \centering
%     \hspace*{-0.6cm}
%     \includegraphics[width = 7in]{plot_kNN_bins_10percent.pdf}
%     \vspace{-0.3cm}
%     \caption{The full target mock $k$NN-CDF($r$). We use $k = 1,2,3,..., 10$ for this analysis. For each $k$, we sample the CDF at 50 linearly spaced scales between $r_\mathrm{min} = 0.1h^{-1}$Mpc and $r_\mathrm{max}=20h^{-1}$Mpc. We further remove scales where the CDF is less than 0.1 or greater than 0.9. The blue curves show the full CDF, whereas the orange points mark the ``bins'' we use as our target data vector.}
%     \label{fig:$k$NN_target}
% \end{figure*}

\section{HOD recovery on mock galaxy lightcones}
\label{sec:recovery}

In this section, we perform a validation test on our lightcone-based forward model by recovering the underlying HOD parameters of a mock galaxy catalog mimicking realistic observations. The purpose is largely to show that parameter inference with lightcone-based full forward models are computational tractable and that such a routine can accurately recover the parameters of interest despite the added layers of model complexity. 

\subsection{Mock data setup}
\label{subsec:mock}

To construct the target mock galaxy catalog, we start with 20 lightcones at Planck cosmology but with different realisations. In the \textsc{AbacusSummit} suite, these lightcones are generated from the phase 005-024 boxes. The other 5 phases (000-004) are reserved for model evaluations. For each of the 20 lightcones, we apply a fiducial HOD whose baseline parameter values are $\log M_\mathrm{cut} = 12.8$, $\log M_1 = 13.9$, $\sigma = 0.3$, $\alpha = 1.0$, $\kappa = 0.3$, and an completeness parameter $\mathrm{ic} = 0.41$. Then we follow the exact steps described in Section~\ref{sec:model} and propagate each of the 20 lightcone catalogs through the CMASS redshift-dependent density filter $n(z)$, the survey window function, and survey masks. Then we measure the desired summary statistics on each of the 20 mocks, and compute the average as the final target statistics. 

\red{The HOD parameters are picked to roughly match that of the CMASS sample \citep{2021bYuan, 2016Rodriguez, 2015Kwan}. These parameters correspond to a satellite fraction of $14\%$ and a number density of $3\times 10^{-4}h^{3}$Mpc$^{-3}$. The average halo mass of the sample is $2\times 10^{13}h^{-1}M_\odot$. For the subsequent analyses, we fix $\sigma$ and $\kappa$ and only vary $\log M_\mathrm{cut}$, $\log M_1$, $\alpha$, and $\mathrm{ic}$. $\kappa$ controls the cut-off mass for satellite galaxies and do not significantly affect clustering for a CMASS LRG-like sample. $\sigma$ has a strong degeneracy with $\log M_\mathrm{cut}$ as they both control the halo mass at which the central galaxy occupation turns off. By fixing $\sigma$, we remove this degeneracy and thus simplify the parameter interpretation and shorten the sampling runs. We will free these parameters in a final analysis of the data. }

To construct the $k$NN($r$) mock data vector, we use the first 10 orders, $k = 1,2,3,..., 10$. For each $k$, we sample the CDF at 50 linearly spaced scales between $r_\mathrm{min} = 0.1h^{-1}$Mpc and $r_\mathrm{max}=20h^{-1}$Mpc. We further remove scales where the CDF is less than 0.1 or greater than 0.9 as these points tend to highly covariant and lead to very poorly behaved covariance matrices. These points also do not contribute much physical information as they are noisey and close to the constrained ends of the CDF. As a result, we end up with 219 points across 10 $k$ values. We illustrate this ``binning'' scheme in Figure~\ref{fig:knn_target}, where the colored markers showcase the full peaked CDFs (pCDFs) and the 219 points we retain. The peaked CDF is adopted for visualisation purposes and is simply defined as $\mathrm{pCDF} = \mathrm{min}(\mathrm{CDF}, 1-\mathrm{CDF})$. We put quotation marks around the word ``binning'' to highlight the fact that we are not in fact integrating the CDF into bins, but simply sampling the CDF at a set of scales. 

\red{Observing the target $k$NN-CDF data vector, we see that we do not utilise scales below $\sim 4h^{-1}$Mpc. Because we expect much of the galaxy--halo connection information is encoded in scales at around or below $\sim 1h^{-1}$Mpc, this current $k$NN setup is likely not optimal for galaxy--halo connection science. To probe smaller scales at fixed galaxy number density, we need a higher density of random query points, which significantly impacts computational performance and can quickly overwhelm the memory. One can also resort to optimal sampling techniques such as the one proposed in Appendix~A of \citet{2022Garrison}. Alternatively, this also means that $k$NN analyses would strongly benefit from higher density samples. Figure~\ref{fig:knn_target} also shows that higher $k$s probe larger scales. This also suggests that including higher $k$s likely has diminishing returns, at least in terms of galaxy--halo connection analyses. }

\begin{figure}
    \centering
    \hspace*{-0.6cm}
    \includegraphics[width = 3.7in]{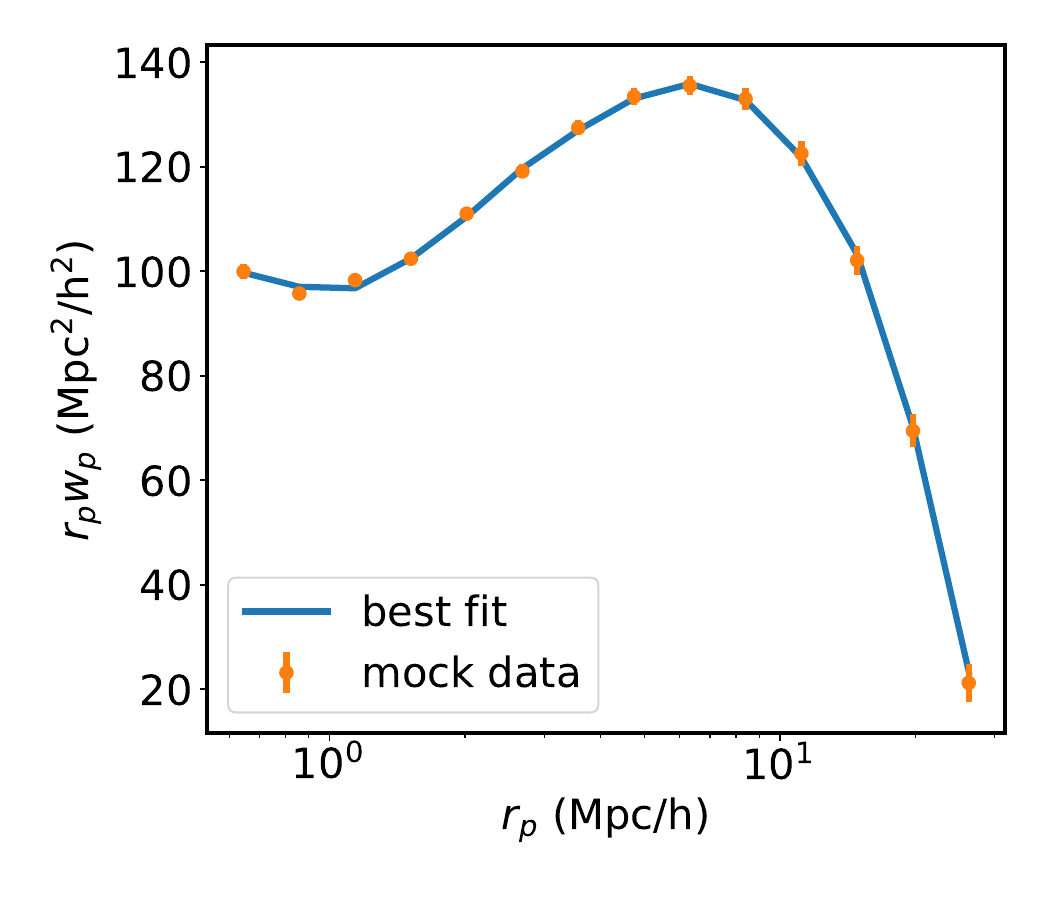}
    \vspace{-0.5cm}
    \caption{The yellow points showcase the target projected 2PCF $w_p$ and its error bars. The $x$ and $y$ axes denote the transverse separation bins $0.5h^{-1}\mathrm{Mpc} < r_p < 30h^{-1}\mathrm{Mpc}$. We plot $r_pw_p$ for visualisation. The blue line represents the best fit we obtain in section~\ref{subsec:recovery}.}
    \label{fig:2pcf_target}
\end{figure}

Similar to the $k$NN-CDF, we compute the projected 2PCF over the 20 fully forward modeled lightcone catalogs and compute the average as our mock target data vector. Specifically, we choose 14 logarithmic bins between 0.5$h^{-1}$Mpc and 30$h^{-1}$Mpc in the transverse separation $r_p$, and a $\pi_\mathrm{max}=30h^{-1}$Mpc. We set the smallest projected scale to match the minimum scale at which our redshift recovery method works well. The target projected 2PCF is visualised with the orange markers in Figure~\ref{fig:2pcf_target}. The error bars represent the expected sample variance in a CMASS volume. 

\begin{figure*}
    \centering
    \hspace*{-0.6cm}
    \includegraphics[width = 6.6in]{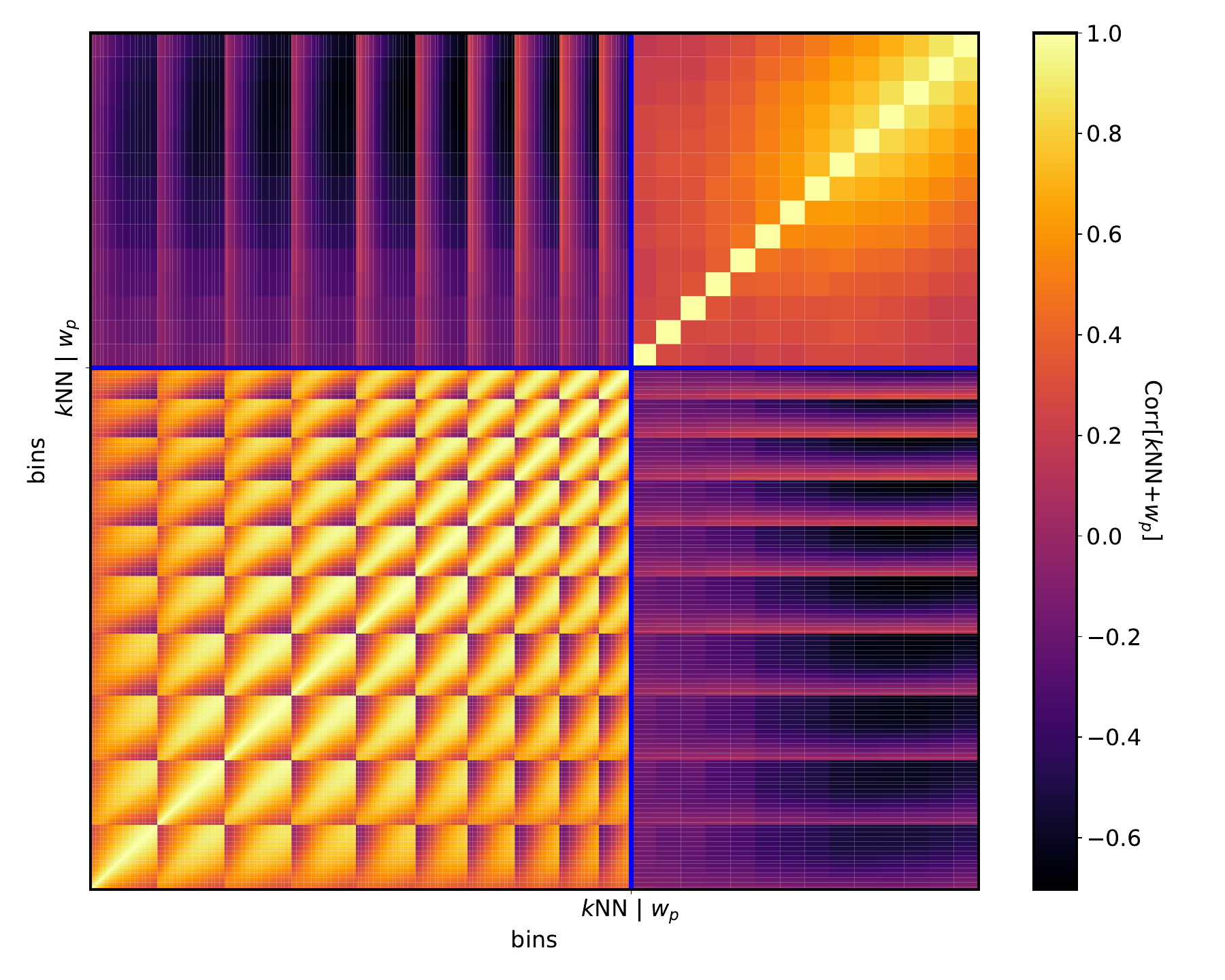}
    \vspace{-0.3cm}
    \caption{\red{The mock joint correlation matrix of the $k$NN-CDF and the projected 2PCF $w_p$. The $x$ and $y$ axes denote bins of the $k$NN-CDF and $w_p$, in that order. The first 219 bins correspond to the flattened $k$NN bins selected across $k = 1,2,3,..., 10$. These bins are ordered such that the blocks represent increasing $k$s, and the bins within each block represent increasing $r$s. The last 14 bins denote the transverse separation bins for $w_p$ between $0.5h^{-1}\mathrm{Mpc} < r_p < 30h^{-1}\mathrm{Mpc}$. We inflate the bin size of these bins for visual clarity. We label the separation of the $k$NN-CDF block and the $w_p$ block with blue lines and axis labels. }}
    \label{fig:cov_kNN}
\end{figure*}

To generate the covariance matrix for likelihood evaluations, we utilise the 1800 \textsc{AbacusSummit} covariance boxes, each of volume $(500h^{-1}$Mpc$)^3$. We apply the fiducial HOD to every box and then calculate the summary statistics, without applying the additional layers of systematics. We then calculate the covariance matrix, which we re-scale to match the CMASS volume. We note that this covariance matrix likely underestimates the actual uncertainties because it only accounts for sample variance and not any of the systematic effects. However, for the purpose of a mock test, we just need a well determined (high signal-to-noise) covariance matrix that is representative of the real covariance structure of the summary statistics. We showcase the joint correlation matrices for the $k$NN-CDF and the 2PCF in Figure~\ref{fig:cov_kNN}. The correlation matrix is simply the covariance matrix normalised by its diagonal elements. 

% \begin{figure}
%     \centering
%     \hspace*{-0.6cm}
%     \includegraphics[width = 3.7in]{plot_wp_cov.pdf}
%     \vspace{-0.3cm}
%     \caption{The mock correlation matrix of the projected 2PCF $w_p$. The $x$ and $y$ axes denote the transverse separation bins $0.5h^{-1}\mathrm{Mpc} < r_p < 30h^{-1}\mathrm{Mpc}$. There is significant off-diagonal power at larger scales, suggesting that the variance on larger scales is dominated by cosmic variance whereas the smaller scales are dominated by shot noise.}
%     \label{fig:cov_wp}
% \end{figure}

The $k$NN-CDF only block shows strong off-diagonal terms. The covariance within each $k$ block along the diagonal is expected as the cumulative distribution function is covariant by definition. The covariance between different $k$ values is also expected as the difference between the different $k$ values results also makes sense as the $k$NN-CDF of order $k$ is closely related to a sum of the counts-in-sphere statistics up to order $k-1$ (Equation~\ref{equ:cdf}). The 2PCF-only block shows significantly less off-diagonal power, especially at small transverse scales, where shot noise dominates. At larger transverse scales, sample variance becomes more important and the bins begin to be correlated. There is moderate cross-correlation between $k$NN-CDF and $w_p$, particularly between small $r$ bins in the $k$NN-CDF and $w_p$.

\subsection{Likelihood model}

\begin{figure*}
    \centering
    \hspace*{-0.6cm}
    \includegraphics[width = 7in]{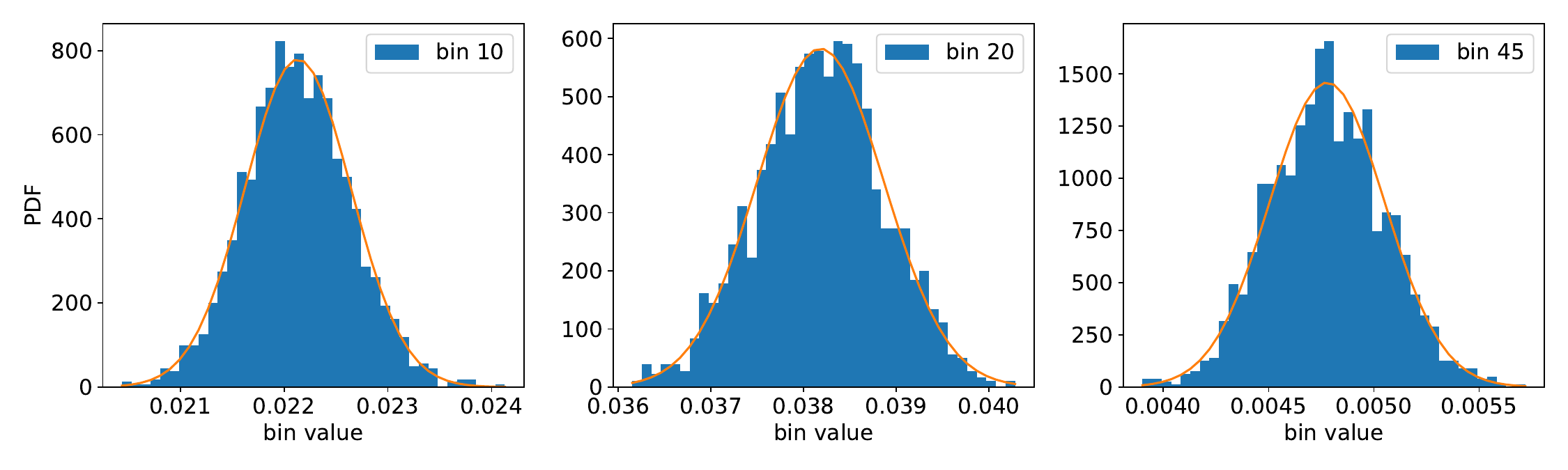}
    \vspace{-0.3cm}
    \caption{The PDFs of 3 arbitrarily chosen $k$NN-CDF bins across 1800 realisations are shown in blue. The orange curves show the Gaussian distribution with the same mean and standard deviation. The $k$NN-CDF does not show any significant non-gaussianity. }
    \label{fig:gaussian}
\end{figure*}

To recover the underlying HOD parameters from the summary statistics computed on the target mocks, we utilise the 5 remaining lightcones, phase 000-004. For each model evaluation, we propose a set of HOD parameters from a flat prior, populate the 5 lightcones with the proposed HOD, and then apply the systematics effects, including redshift selection, survey window and masks. Finally, we compute the summary statistics averaged over the 5 lightcones, which we compare with the target summary statistics and calculate likelihoods using the aforementioned covariance matrix. For this analysis, we adopt a Gaussian likelihood function that accounts for both the desired summary statistics and also the average density. Specifically, 
\begin{align}
    \log L = &  \frac{1}{2}(x_\mathrm{proposed} - x_\mathrm{target})^T \boldsymbol{C}^{-1}(x_\mathrm{proposed} - x_\mathrm{target}) \nonumber\\
    & + \frac{1}{2}\frac{(\bar{n}-\bar{n}_\mathrm{target})^2}{\sigma_n^2}
\end{align}
where $x$ is the desired summary statistic, $\boldsymbol{C}$ is the covariance matrix, and $\bar{n}$ is the mean number density. $\sigma_n$ is the uncertainty on the measured mean number density. For a CMASS-like sample, we quote $\sigma_n = 5\%$ \citep[][]{2022Yuan, 2015aGuo}.
Here we have assumed a Gaussian likelihood, which is known to be the case for the 2PCF. However, for $k$NN-CDF, we test its Gaussianity with the 1800 realisations we have available through the small boxes. Figure~\ref{fig:gaussian} shows the distribution of 3 arbitrary $k$NN-CDF bins across the 1800 realisations, and we do not see any significant non-gaussianity. 

\subsection{Emulator}

Typically, to sample an HOD parameter space until convergence, approximately $10^5-10^6$ likelihood evaluations are required. For this analysis, we use the \textsc{dynesty} nested sampler \citep{2018Speagle, 2019Speagle} as it can sample the posterior space more efficiently than an Markov Chain Monte Carlo sampler. However, given that each forward model evaluation takes approximately 1.5 seconds per lightcone on our machine, and we are evaluating 5 lightcones per likelihood call, a 1,000,000 call chain would take more than 80 days. Thus, we adopt an emulator scheme to speed up the likelihood evaluation. 

In cosmology, an emulator refers to a scheme where one interpolates sparse likelihood evaluations with a smooth parametrised model, also referred to as the surrogate model or just the emulator. By training such an emulator model, the idea is to replace the expensive likelihood calls with the much cheaper emulator model calls, thus enabling a much faster sampling at the cost of introducing additional errors in the model training. Such emulation schemes have become increasingly popular with the advent of fast yet flexible machine learning models such as neural nets and Gaussian processes, with a series of successful cosmology applications in recent years \citep[e.g. ][]{2009Heitmann, 2010Lawrence, 2014Heitmann, 2019Zhai, 2022Zhai, 2021Lange, 2021Kobayashi, 2022bYuan}.

For this analysis, we construct a fully connected neural network as our surrogate model, taking in HOD parameters and outputting the fully forward-modeled summary statistics. For the $k$NN-CDF, we adopt a network of 3 layers as our fiducial model, with 200 nodes in each layer and Randomised Leaky Rectified Linear Units (RReLU) activation. We train the network with the Adam optimiser and a mean squared loss function, where we use the diagonal terms of the mock-based covariance matrix as bin weights. For the 2PCF, we find a 2 layer network to work best, with 100 nodes in each layer and RReLU activation. For testing and validation, we set aside $10\%$ of the training sample as the test set and another $10\%$ as the validation set. For training, we follow a mini-batch routine, where the training set is divided into 100 equal batches, which are then passed the optimiser one at a time. 

To generate the training set, we follow the hybrid MCMC+emulator approach first implemented in \citet{2022bYuan}. Since we know the target data vector and the covariance matrix, we can directly sample the likelihood surface with a set of MCMC (Markov Chain Monte Carlo) chains. However, instead of running the chains till convergence, we stop the chain once a certain number of likelihood calls has been reached, as limited by the compute time available. We then use these samples generated by the MCMC as the training set for the emulator. Compared to the standard method where the training set is generated with a space-filling sampling of the parameter space, such as a Latin Hypercube, this method allows for a significantly tighter prior region, resulting in higher density of training samples and thus smaller emulator errors. In this approach, we can also think of the emulator step as continuing the MCMC chain, except with a surrogate likelihood model that is orders of magnitudes faster to calculate. However, with a smaller training range, we also need to make sure that the training is robust against biases towards the mean. We note that similar iterative sampling-emulation ideas were also discussed in \citet{2020Pellejero}.

To fit the target $k$NN-CDF, we first run an MCMC chain against the target $k$NN-CDF stopped at 100,000 likelihood calls to generate 100,000 training points. Then we impose a likelihood cut to select the 40,000 training points with the highest likelihoods. This sample then undergoes the 80/10/10 training/validation/test split. The resulting training set is then used to train the neural network, and the validation set is used to check for over-fitting during the training. When the training converges, we test the best-fit model on the test set. We present the following test results. 

\begin{figure}
    \centering
    \hspace*{-0.6cm}
    \includegraphics[width = 3.9in]{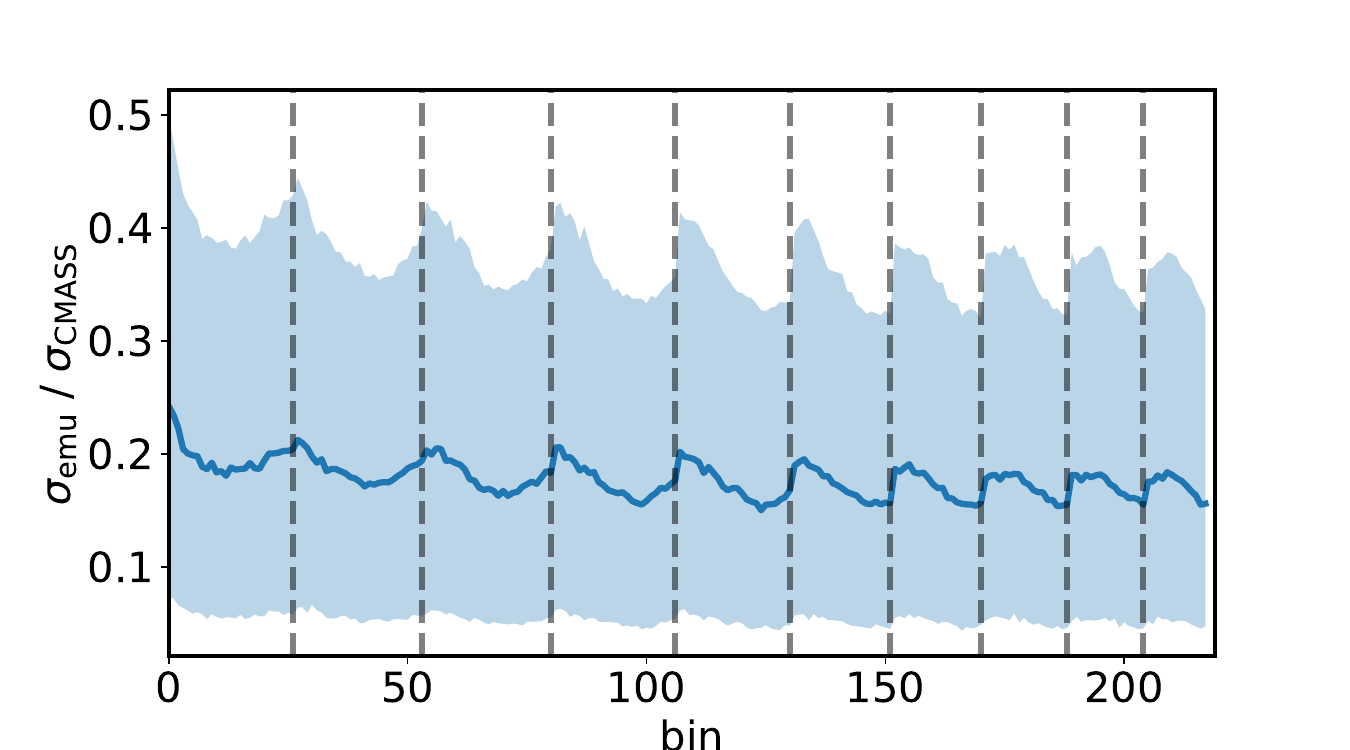}
    \vspace{-0.3cm}
    \caption{The best-fit emulator's absolute outsample error relative to CMASS sample variance. The $x$ axis denotes the bin number, with each $k$ order separated by the dashed vertical lines. The bin number increases with scale within each $k$ section. The blue line denotes the median absolute error, whereas the shaded region denotes the extent of the $1\sigma$ region. Clearly, the emulator errors are well within the limit of CMASS sample variance. }
    \label{fig:emu_percentile}
\end{figure}

Figure~\ref{fig:emu_percentile} presents the outsample error of $k$NN($r$) as a fraction of the expected error due to sample variance in a CMASS volume. The blue line denotes the median absolute error whereas the shaded region denotes the extent of the $1\sigma$ region. clearly, the emulator error is significantly smaller than the expected CMASS sample variance. Averaging over all the tests and bins, we get a representative value of 0.28, which is the mean emulator error as a fraction of the expected sample variance error. 

\begin{figure*}
    \centering
    \hspace*{-0.6cm}
    \includegraphics[width = 6.5in]{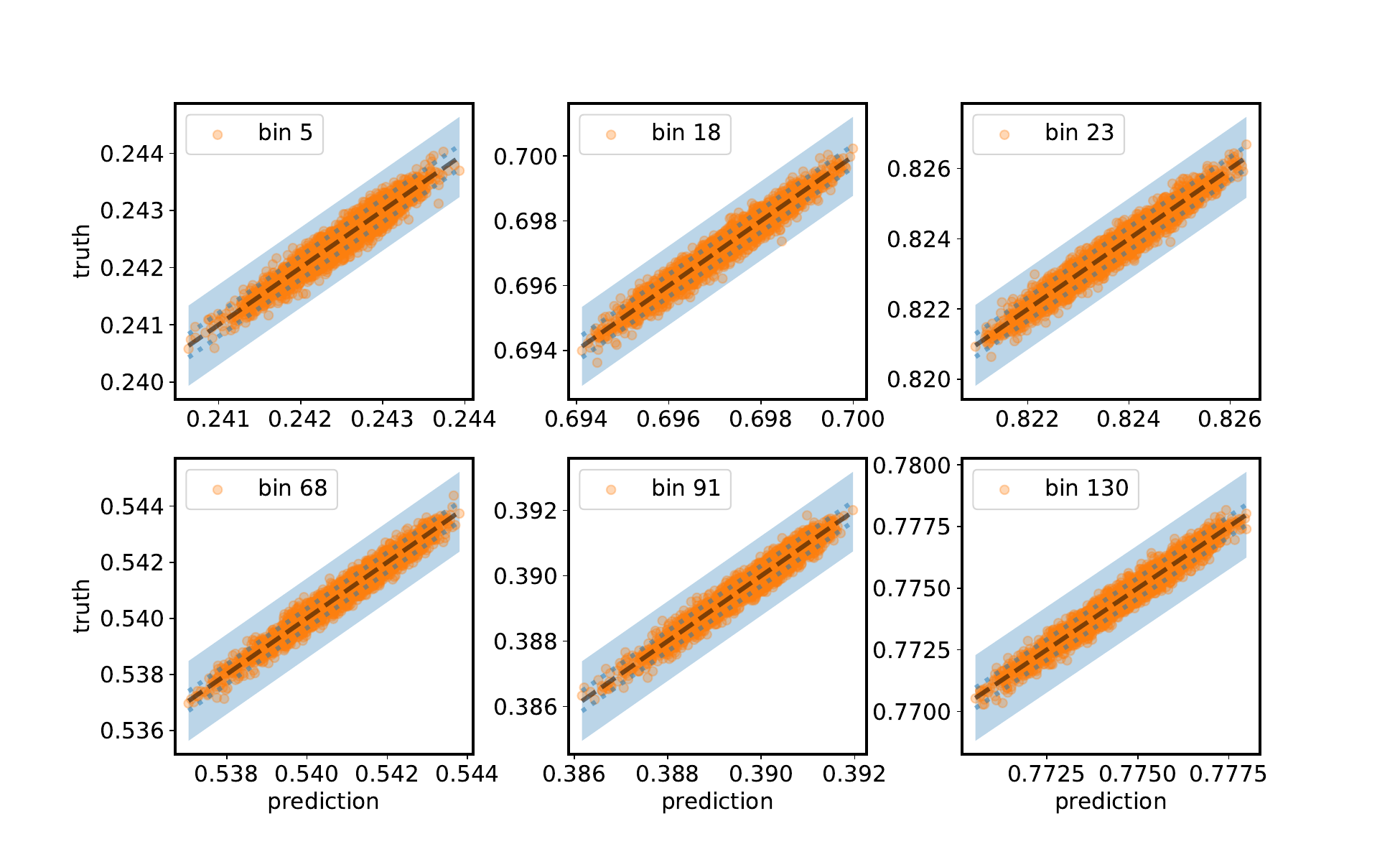}
    \vspace{-0.6cm}
    \caption{The true $k$NN-CDF bin values versus the predicted values from the emulator in a 6 randomly selected bins (orange points). The dashed black line denotes equality for reference. The blue shaded region showcase the CMASS sample variance (1$\sigma$). The green dotted lines denote the standard deviation of the scatter. We do not see any significant bias in the prediction that depends on the true values. The scatter is well within the expected sample variance.}
    \label{fig:emu_bias}
\end{figure*}

Figure~\ref{fig:emu_bias} presents the scatter plot of the true values and the emulator-predicted values for a few randomly selected bins. Again, we see that the prediction error is well within the expected CMASS sample variance, which is shown by the blue band. The plot also shows that there is no significant bias towards the mean in the emulator prediction. Thus, we deem the best-fit neural net model to be unbiased and sufficiently accurate to replace the original likelihood calculations without introducing significant additional errors, at least within the training range. 

We follow the exact same procedure for the 2PCF. We conduct tests to ensure that the resulting emulator error is subdominant compared to the data error, and that the emulator predictions are not biased towards the mean of the training range. The resulting mean emulator errors as a fraction of the data error are also around $28\%$ for $k$NN($r$) and $30\%$ for the 2PCF. We do not show the figures for the 2PCF for brevity.

\subsection{Parameter recovery}
\label{subsec:recovery}
Having trained and tested the emulators for the $k$NNs and the 2PCF, we can test the constraints of these two summary statistics by sampling the parameter posteriors given the mock data vectors (Figure~\ref{fig:knn_target} and Figure~\ref{fig:2pcf_target}) and the mock covariance matrices (Figure~\ref{fig:cov_kNN}. \red{For this analysis, we run three chains: one with just $k$NN-CDF, one with just $w_p$, and one with both data vectors. } 

For faster sampling, we use the \textsc{dynesty} nested sampler \citep{2018Speagle, 2019Speagle}. We also impose flat priors bounded with an ellipsoid for all parameters. The ellipsoid is constructed as the minimum-volume ellipsoid that envelopes all training points. We initiate each nested sampling chain with 2000 live points and a stopping criterion of $d\log\mathcal{Z} = 0.01$, where $\mathcal{Z}$ is the evidence. As expected, we achieve excellent fits for both summary statistics, with best-fit $\chi^2$/d.o.f $<1$. 

\begin{figure}
    \centering
    \hspace*{-0.6cm}
    \includegraphics[width = 3.9in]{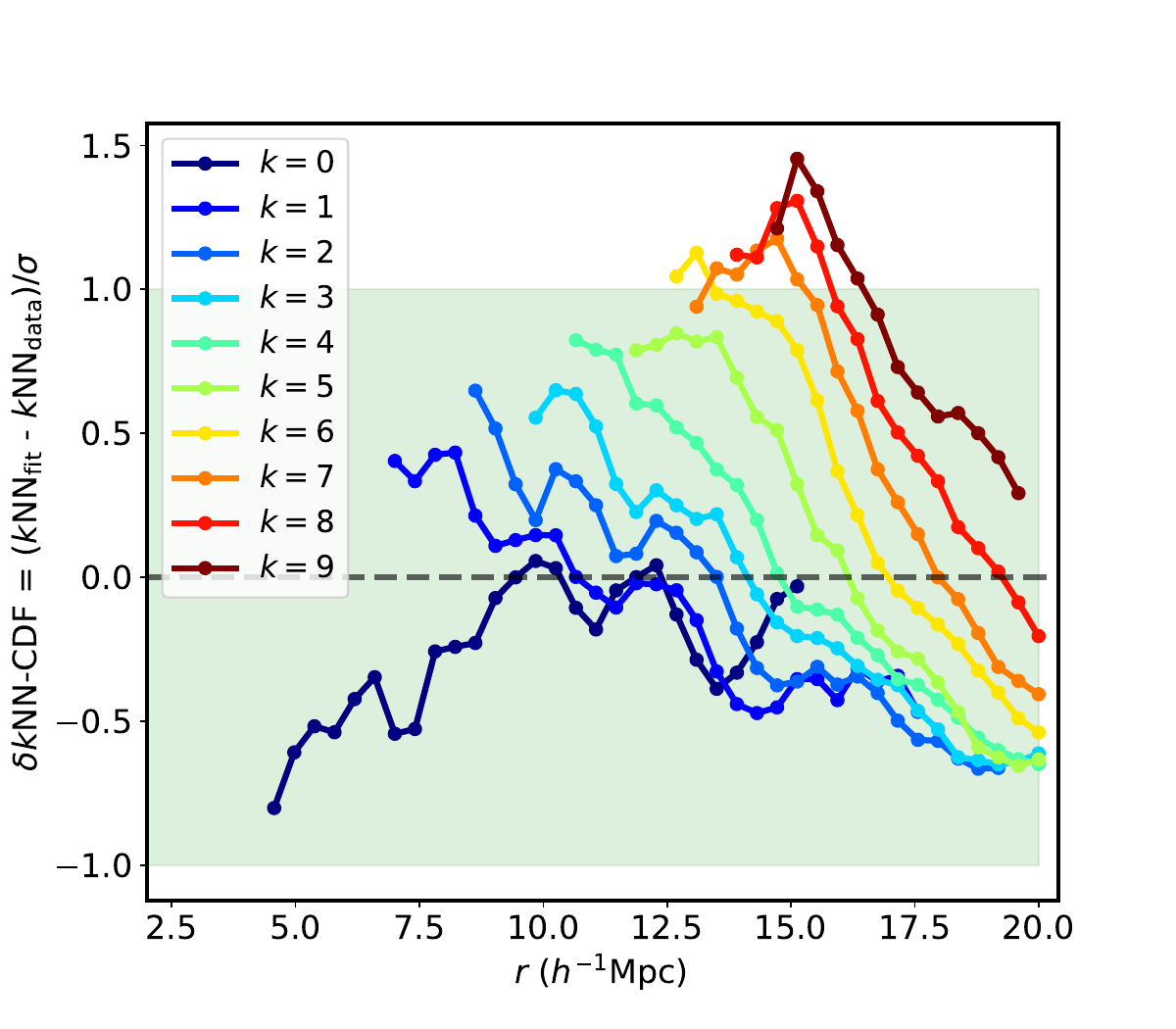}
    \vspace{-0.4cm}
    \caption{\red{The difference between the best-fit $k$NN-CDF and the mock data vector. The $y$-axis shows the relative difference between the best fit and the mock data, normalised by the CMASS error bar. The green band indicate the $1\sigma$ region. Different colors correspond to different $k$s.}}
    \label{fig:knn_fit}
\end{figure}

\red{Figure~\ref{fig:knn_target} and Figure~\ref{fig:2pcf_target} showcase the best fit predictions compared to the target data vectors. We achieve good fits visually in both cases. However, because the error bars on the $k$NN-CDF are tiny on an absolute scale, we explicitly show the difference between the best fit and the target data vector in Figure~\ref{fig:knn_fit}, normalised by the CMASS error bars. We see that for most $k$s, the best-fit residual falls well within the $1\sigma$ CMASS error. At the highest $k$s, there is a slightly larger residual at smaller scales. }

\begin{figure*}
    \centering
    \hspace*{-0.6cm}
    \includegraphics[width = 6in]{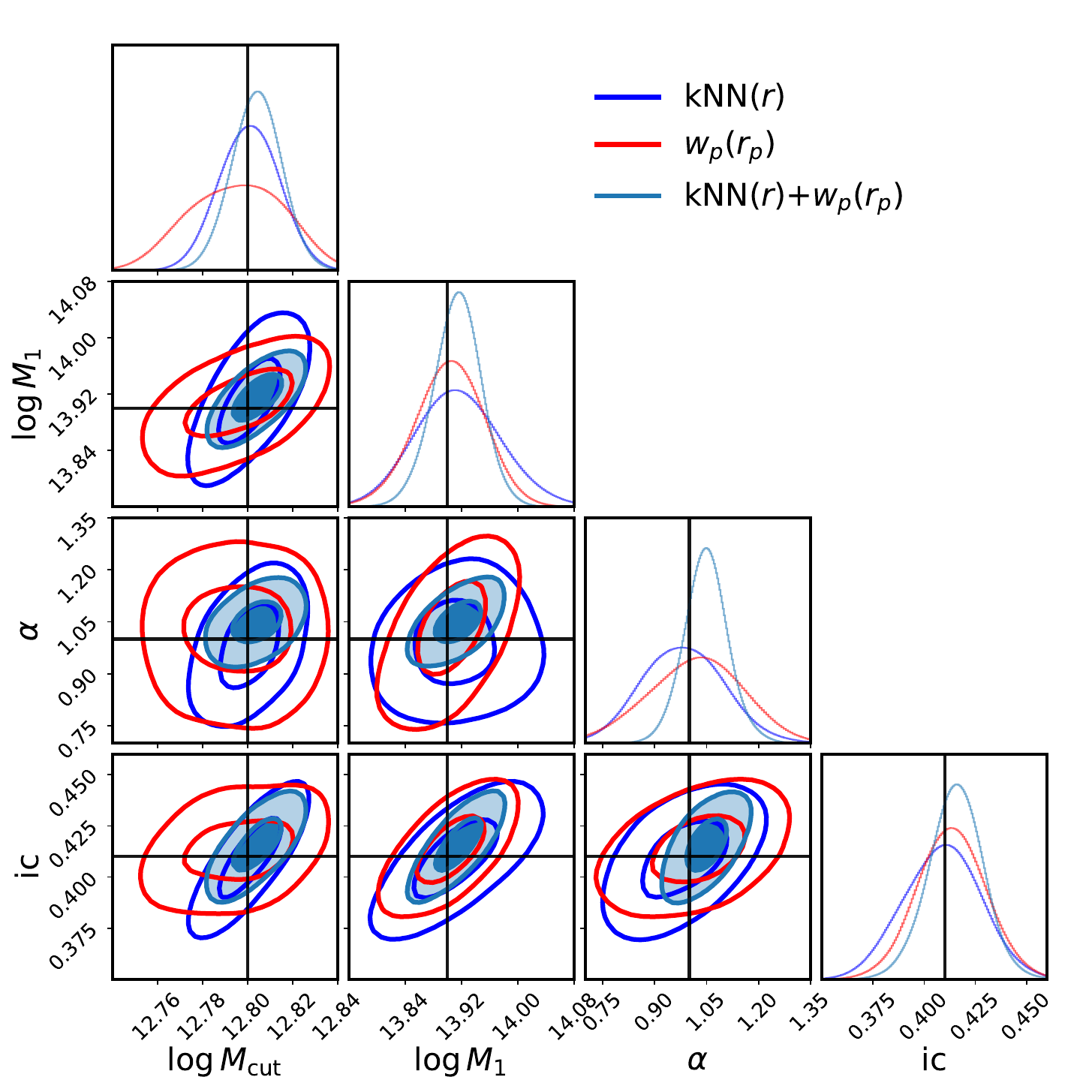}
    \vspace{-0.3cm}
    \caption{\red{The HOD posterior as recovered by the projected 2PCF $w_p$ and the $k$NN($r$). The black lines denote the truth values, whereas the contours denote the 1 and 2$\sigma$ constraints. The green contours showcase the joint constraints of the two data vectors.} }
    \label{fig:corner}
\end{figure*}

We present the resulting 2D marginalised posteriors in Figure~\ref{fig:corner} and also summarise the 1D marginalised constraints in Table~\ref{tab:constraints}. The blue and red contours denote the 1-2$\sigma$ constraints from the $k$NNs and $w_p$, respectively. The green contours showcase the joint constraints with $k$NN+$w_p$. The black lines show the truth values. %The titles above the 1D marginalised posteriors list the 95$\%$ or $2\sigma$ constraints for each parameter. 
The main conclusion is that our full forward model approach can obtain unbiased recoveries of all model parameters despite the layers of observational systematics. This is significant for analyses with beyond-2PCF statistics on non-linear scales as our full forward approach does not rely on any specific summary statistics. We have demonstrated that a full forward model for any summary statistic that accounts for the full range of observational systematics is computationally viable and should yield unbiased constraints. 

\begin{table*}
\centering
\begin{tabular}{lllccc}
\hline
Parameter    & Meaning & truth & $w_p$ post. ($95\%$C.L.) & $k$NN post. ($95\%$C.L.) & joint post. ($95\%$C.L.)\\
\hline
$\log_{10}{M_\mathrm{cut}}$  & The typical mass scale to host a central  & 12.8 & $12.79\pm 0.04$ & $12.80\pm 0.02$ & $12.80\pm 0.02$\\
$\log_{10}{M_{1}}$  & The typical mass scale for halos to host one satellite  & 13.9 & $13.90\pm 0.09$ & $13.91\pm 0.11$ & $13.91\pm 0.06$\\
$\alpha$  & The power-law index for satellites & 1.0 & $1.0\pm 0.2$ & $1.0\pm 0.2$ & $1.05\pm 0.11$\\
$\mathrm{ic}$  & The incompleteness parameter & 0.41 & $0.41\pm 0.03$ & $0.41\pm 0.03$ & $0.41\pm 0.03$\\

\hline
\end{tabular}

\caption{\red{The marginalised posterior constraints of the 4 HOD model parameters recovered from the projected 2PCF $w_p$ and the $k$NN-CDFs. The last column showcase the joint constraints. The 2D marginalized constrants are visualized in Figure~\ref{fig:corner}.}}
\label{tab:constraints}
\end{table*}

Comparing the $k$NN and 2PCF constraints, we see that the $k$NNs derive competitive constraints compared to the 2PCF. Specifically, the $k$NNs yield stronger constraints on $\log M_\mathrm{cut}$ while deriving slightly weaker constraints on $\log M_1$. 
% while the two statistics derive essentially the same constraints on $\log M_\mathrm{cut}$, the $k$NNs derive considerably weaker constraints on $\log M_1$ and $\alpha$. 
The fact that the $k$NNs derive stronger constraints on $\log M_\mathrm{cut}$ shows the promise of $k$NNs in a cosmology analysis. This is because $\log M_\mathrm{cut}$ is the only parameter controlling the occupation of the centrals since we have fixed $\sigma$. Given that the satellite fraction is small, the central occupation largely controls the 2-halo term in the clustering and thus the linear bias. Thus, strong constraints on $\log M_\mathrm{cut}$ translates to strong constraints on the amplitude of the linear power spectrum and the growth of structure. This is consistent with the Fisher analysis results of \citet{2021Banerjee}.

We do, however, expect that the 2PCF would be more constraining in the satellite occupation parameters than the $k$NNs. On the one hand, the $k$NNs measure the counts of galaxies around randomly selected query points. In a clustered data set, the randomly selected query points will necessarily mostly sample the under-dense regions more than the over-dense regions. The 2PCF, on the other hand, measure the data-data counts, thus it necessarily samples mostly the over-dense regions, boosting the sensitivity to 1-halo scaling clustering $r < 1h^{-1}$Mpc. And because we expect the 2PCF to be more sensitive to 1-halo scale clustering, we expect it to have stronger constraints on satellite occupation parameters $\log M_1$ and $\alpha$. It is surprising to find the $k$NNs to be competitive even in the satellite occupation parameters. This might be due to the fact that we include higher $k$ orders for $k$NNs but only consider the projected 2PCF in this analysis. In our companion paper \citet{2022dYuan}, we develop variations to the standard $k$NN formulation that exceeds the constraining power of the full-shape 2PCF, even on satellite occupation parameters. 

It is also worth noting that this is not a fair comparison since we have not used the $k$NN below $r < 4 h^{-1}$Mpc (see Figure~\ref{fig:knn_target}), whereas we used the 2PCF all the way down to $0.5h^{-1}$Mpc. The cut on the CDF is placed to remove low signal-to-noise points in the $k$NN and is limited by the grid spacing for the query points we used. A denser query point set would increase the signal-to-noise in the $k$NN on small scales but allow us to meaningfully use those scales at the cost of significantly slower likelihood evaluation. In Appendix~\ref{a:rpmin}, we present an alternative comparison of the two statistics by applying a minimum scale cut of $r_{p,\mathrm{min}} = 3h^{-1}$Mpc on the 2PCF, in which case the 2PCF becomes significantly less constraining than the $k$NNs. 

\red{Figure~\ref{fig:corner} also shows significantly stronger joint constraints when the two summary statistics are used simultaneously. This is particularly true for the satellite occupation parameters where we get an approximately factor of 2 improvement in the constraining power. The joint constraints on the central occupation parameter $\log M_\mathrm{cut}$ is not significantly stronger than the $k$NN-CDF constraints, again demonstrating the rich information captured by the $k$NNs. The 1D joint posteriors are again summarised in Table~\ref{tab:constraints}.}

\section{Discussions}
\label{sec:discuss}

% cosmology
A key limitation of this analysis is that we have only tested the recovery of HOD parameters at fixed cosmology. The goal of utilizing the full information of the smaller scales is to learn both about galaxy--halo connection and the underlying cosmology. In order to enable the cosmology dependence of the full forward model, we need simulation lightcones at non-standard cosmologies. The \textsc{Quijote} suite provides lightcones with variable cosmologies, but the resolution is not sufficient for modeling small scales \citep{2020Villaescusa}. \red{A pilot study using this suite in a forward model is described in \cite{2022Hahn}.} Efforts are currently underway to produce high-fidelity lightcones at variable cosmologies from the periodic boxes of the \textsc{AbacusSummit} suite \citep{2022Hadzhiyska}. We reserve the description of these products and the development of cosmology-dependent full forward models for a future paper. 

% hodz
The use of simulation of lightcones also enables direct modeling of redshift-dependence in the bias model, whether it be an HOD($z$) or another galay--halo connection model. This is particularly important given the depth of current and upcoming surveys. For example, with DESI, the LRG sample is expected to span redshift range $0.3 < z < 0.8$ whereas the ELG sample is expected to span $0.6 < z < 1.6$ \citep{2020Zhou, 2020Raichoor}. One clearly expects significantly redshift evolution in the galaxy--halo connection throughout these wide ranges. In a standard approach, one can divide the data into several redshift bins, and analyse each bin with a model template constructed on snapshots of the periodic box at or close to the effective redshift within each bin, such approaches suffer from potential biases due to redshift binning and covariances between different redshift bins that might be hard to account for in a combined analyses. Instead, we propose the use of our lightcone-based forward model approach where we directly parameterise the redshift evolution in the galaxy bias model and compare data to continuous mocks that span the entire redshift range. We would still need to compute the summary statistics in redshift bins, but as long as we make consistent choices between the lightcone and the data, we would not suffer from any biases due to redshift binning. The model covariance between different redshifts are also naturally accounted for. A description of redshift-dependent HOD implementaion in \textsc{AbacusHOD} is presented in Appendix~\ref{a:hodz}.

% number density
\red{As mentioned in section~\ref{subsubsec:knn} and section~\ref{subsec:mock}, $k$NN analyses of galaxy--halo connection would strongly benefit from a higher density galaxy sample which would open up the smaller scales. This is particularly relevant for the DESI Bright Galaxy Survey \citep{2022bHahn}, which will target a magnitude limited galaxy sample at $z < 0.4$. The BGS sample will go significantly fainter but denser than the LRG and ELG sample, and will be a key sample for lensing, high-order statistics, and galaxy--halo connection science. The sample will reach a number density of $3\times 10^{-3}h^3$Mpc$^{-3}$ at $z = 0.4$, i.e. 10 times that of the CMASS sample, and $1\times 10^{-2}h^3$Mpc$^{-3}$ at $z = 0.1$. Such high number density will allow the $k$NN-CDF to probe 1-halo scales at $r < 1h^{-1}$Mpc. We also expect the BGS sample to be highly redshift-dependent. Thus, a lightcone-based forward model combined with our redshift-dependent HOD model would be the natural choice to study the BGS sample. }

% volume
While lightcone-based models hold great promises, generating high-precision lightcones of sufficient volume can be a computational challenge. In this analysis, we average over 5 different lightcones (each covering an octant of the sky) at fixed cosmology to achieve the desired volume in the model and ensure the model sample variance is subdominant compared to other sources of uncertainties. With current lightcone implementations such as the one used in \citet{2022Hadzhiyska}, one can only generate one independent lightcone per simulation box. Thus for a cosmology+HOD analysis, one would require repeat simulations per cosmology or a single box that is significantly larger then what is currently available. 

However, we might not need repeat lightcones to reduce sample variance after all. At small scales, sample variance is subdominant to other modeling uncertainties and observational systematics, in which case a single lightcone would be sufficient. At larger scales, there are sample variance suppression techniques such as the one developed in \citet{2022Kokron}, which uses fast repeat runs of Zel'dovich codes to achieve orders of magnitude reduction to sample variance on large scales. In principle, supplementing lightcone-based predictions with this technique should result in sufficiently precise model templates across a wide range scales. 

% photoz

\section{Conclusions}
\label{sec:conclude}

In this paper, we constructed and tested a full forward modeling pipeline for galaxy clustering statistics that utilises high-fidelity simulation lightcones and account for the full range of observational systematics. We demonstrated that one can recover unbiased model constraints using our forward modeling pipeline with the standard 2PCF and the novel $k$-th nearest neighbor statistics (sensitive to correlation functions of all orders) on non-linear scales. While we used two summary statistics as examples, our pipeline is agnostic to the statistics used and we fully expect the technique is broadly applicable to other novel summary statistics. This is significant for analyses of upcoming cosmological surveys where the use of novel summary statistics and of non-linear scales are necessary to extract the full information content of the vast datasets. 

As a part of the pipeline, we also introduced a novel treatment of fibre collision effects. Specifically, we use the measured clustering to probabilistically recover the redshifts of fibre collided targets. We tested this technique on both the projected 2PCF and the $k$NNs and found good redshift recovery compared to other systematics budget. This method is promising as it does not appeal to any specific properties of summary statistics and is in principle applicable to a wide array of beyond-2PCF analyses. We propose additional testing of the method and potential enhancements by leveraging additional information, such as combining with photometric redshift inferences. 

By testing our pipeline with two different summary statistics, we also produced a realistic comparison of the $k$-th nearest neighbor statistics to the standard 2PCF. We showed that the $k$NNs derive stronger constraints on the linear bias, and the two statistics are similarly informative of the properties of satellite galaxies. We explore additional variations to the $k$NN formalism in the companion paper \citet{2022dYuan}.

\section*{Acknowledgements}

We would like to thank Risa Wechsler, Arka Banerjee, Ashley Ross, Sebastian Wagner-Carena, Philip Mansfield for useful feedback and suggestions in various stages of this analysis.

This work was supported by U.S. Department of Energy through grant DE-SC0013718 and 
under DE-AC02-76SF00515 to SLAC National Accelerator Laboratory. 

This work used resources of the National Energy Research Scientific Computing Center (NERSC), a U.S. Department of Energy Office of Science User Facility located at Lawrence Berkeley National Laboratory, operated under Contract No. DE-AC02-05CH11231.

The {\sc AbacusSummit} simulations were conducted at the Oak Ridge Leadership Computing Facility, which is a DOE Office of Science User Facility supported under Contract DE-AC05-00OR22725, through support from projects AST135 and AST145, the latter through the Department of Energy ALCC program.

%%%%%%%%%%%%%%%%%%%%%%%%%%%%%%%%%%%%%%%%%%%%%%%%%%
\section*{Data Availability}

The simulation data are available at \url{https://abacussummit.readthedocs.io/en/latest/}. The \ahod\ code package is publicly available as a part of the \textsc{abacusutils} package at \url{http://https://github.com/abacusorg/abacusutils}. Example usage can be found at \url{https://abacusutils.readthedocs.io/en/latest/hod.html}.



\bibliographystyle{mnras}
\bibliography{biblio} % if your bibtex file is called example.bib

%\begin{thebibliography}{99}
%\end{thebibliography}

%%%%%%%%%%%%%%%%%%%%%%%%%%%%%%%%%%%%%%%%%%%%%%%%%%

%%%%%%%%%%%%%%%%% APPENDICES %%%%%%%%%%%%%%%%%%%%%

%%%%%%%%%%%%%%%%%%%%%%%%%%%%%%%%%%%%%%%%%%%%%%%%%%
\appendix

\section{redshift-dependent HOD}
\label{a:hodz}
\red{
The \ahod package also enables populating lightcones with redshift-dependent HODs. This is particularly powerful for current and future deep surveys such as DESI and PFS that observe galaxies over a large redshift range. In such samples, modeling redshift evolution not only helps our understanding of galaxy evolution, but also is necessary to avoid biasing the cosmology constraints and thus leverage the full statistical power of the data. }

In a first implementation, we add two additional free parameters to the model, $\mu_{\mathrm{cut}, p}$ and $\mu_{1, p}$, where we define $\mu \equiv \log M $. These two parameters are the first derivative of $\log M_{\rm cut} $ and $\log M_1$ against the scale factor, respectively. Thus, the modified parameters in the \textsc{AbacusHOD} model take the following form for a given choice of a reference redshift, $z_{\rm pivot}$,
\begin{equation}
    \log M_i (z) = \log M_i (z_{\rm pivot}) + \mu_{i, p} \left(\frac{1}{1+z} - \frac{1}{1+z_{\rm pivot}}\right) ,
\end{equation}
where $i = \{{\rm cut}, \ 1\}$. Currently, we opt to make only $M_{\rm cut}$ and $M_1$ redshift-dependent for simplicity. In principle, all HOD parameters can be redshift-dependent and the \ahod\ package can be easily extended to accommodate such complexities. 

\section{Comparing the statistics on equivalent scales}
\label{a:rpmin}
\begin{figure*}
    \centering
    \hspace*{-0.6cm}
    \includegraphics[width = 5.5in]{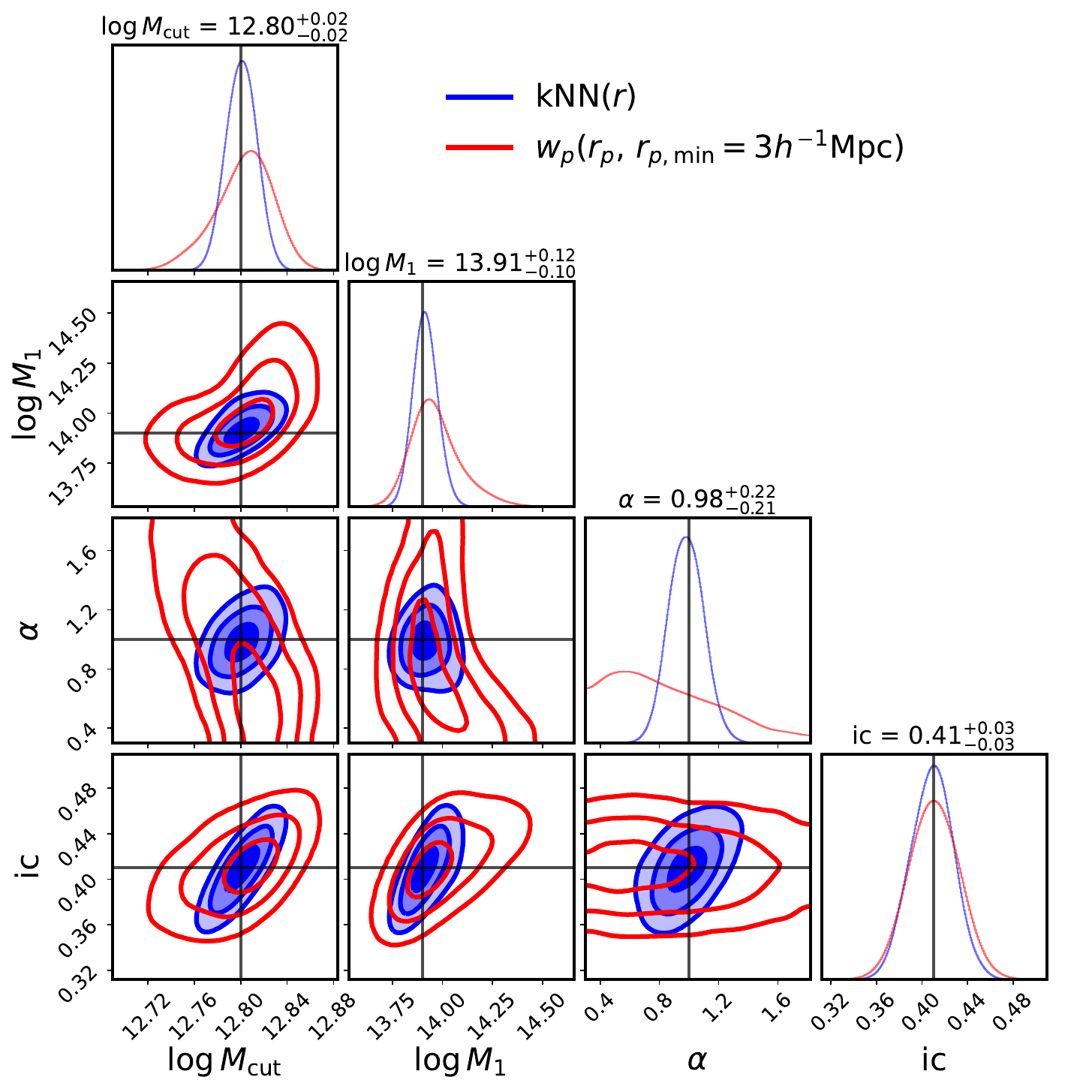}
    \vspace{-0.3cm}
    \caption{The HOD posterior as recovered by the projected 2PCF $w_p$ with $r_{p,\mathrm{min}} = 3h^{-1}$Mpc and the $k$NN($r$). The black lines denote the truth values, whereas the contours denote the 1-3$\sigma$ constraints. The titles on the 1D histograms describe $95\%$ confidence interval of the $k$NN constraints. }
    \label{fig:corner_rpmin}
\end{figure*}

Figure~\ref{fig:corner} appears to show the 2PCF as more informative on the HOD than $k$NNs. However, as we pointed out towards the end of section~\ref{subsec:recovery}, we have only utilised the $k$NNs on scales above $r > 4h^{-1}$Mpc due to the relatively sparse query set, whereas we used the 2PCF all the way down to $0.5h^{-1}$Mpc. Here we facilitate a comparison of the two statistics at equivalent scales, specifically by only using the 2PCF at scales $r_p > 3h^{-1}$Mpc. We follow the exact same procedure as laid out in section~\ref{sec:recovery} and show the resulting constraints in Figure~\ref{fig:corner_rpmin}. 

We see that the constraining power of the projected 2PCF $w_p$ decreases considerably in both the central and satellite HOD parameters. The constraints on the satellite parameter $\alpha$ turns out particularly poor, hitting prior bounds in both directions. The $w_p$ at $r_p > 3h^{-1}$Mpc still derives strong constraints on the mass parameters $\log M_\mathrm{cut}$ and $M_1$, but the constraining power is now significantly inferior to that of the $k$NN. 

Finally, we note that the definition of scales is different between the 2PCF and $k$NNs, with the 2PCF defining the separation between data-data pairs and $k$NNs defining the separation between data-query pairs. The idea of testing on equivalent scales is an attempt to compare the statistics in roughly the same 2-halo regime, but an exactly fair comparison is not possible. We also point to the fact that there are significantly more bins available in the $k$NN than the projected 2PCF. Ultimately, these results highlight the complementarity of the two statistics and the potential information gain by combining the two.

% Don't change these lines
\bsp	% typesetting comment
\label{lastpage}
\end{document}